\documentclass[nofootinbib,superscriptaddress,floatfix,10pt,aps,prd,preprintnumbers]{revtex4-2}

\usepackage{amsmath,amssymb,graphicx,bm,physics,slashed,bbm,multirow,comment}
\usepackage{setspace}
\usepackage{hyperref}
\usepackage{orcidlink}

\makeatletter
\preto\env@matrix{}
\makeatother

\begin{document}

\preprint{COMETA-2026-45}

\title{Spectroscopy of the SU(2) scalar-fermion-gauge system \\ under the influence of the Brout-Englert-Higgs effect}

\author{Axel Maas\,\orcidlink{0000-0002-4621-2151}}
\email{axel.maas@uni-graz.at}
\affiliation{Institute of Physics, NAWI Graz, University of Graz, Universitätsplatz 5, 8010 Graz, Austria}

\author{Sofie Martins\,\orcidlink{0000-0001-5428-8925}}
\email{sofie.martins@uni-graz.at}
\affiliation{Institute of Physics, NAWI Graz, University of Graz, Universitätsplatz 5, 8010 Graz, Austria}

\author{Georg Wieland\,\orcidlink{0009-0003-1777-4563}}
\email{georg.wieland@uni-graz.at}
\affiliation{Institute of Physics, NAWI Graz, University of Graz, Universitätsplatz 5, 8010 Graz, Austria}

\date{\today}

\begin{abstract}
    We investigate the ground state spectrum of a dynamical scalar-fermion-gauge system with an SU(2) gauge group, a scalar doublet, and two degenerate generations of vectorial fermions under the influence of the Brout-Englert-Higgs effect. To this end, we employ lattice methods using the HiRep code. We study the dependence of the system on the (bare) fermion mass and find that the spectrum remains unaffected, up to trivial fermion mass effects. The results are in line with a weakly coupled theory exhibiting the Fr\"ohlich-Morchio-Strocchi mechanism.
\end{abstract}

\maketitle

\section{Introduction}
\label{sec:intro}

Electroweak physics and leptons are tremendously well~\cite{pdg} described by perturbation theory~\cite{Bohm:2001yx}. However, a number of field-theoretical theorems make this, at first sight, surprising~\cite{Maas:2017wzi}. In particular, Elitzur's theorem~\cite{Elitzur:1975im} implies that the instrumental electroweak symmetry breaking can be only a figure of speech: No gauge symmetry can be spontaneously broken, at most it can be hidden. However, electroweak symmetry breaking is instrumental in the construction of the asymptotic states in perturbation theory~\cite{Bohm:2001yx}, allowing them to be identified with the elementary states. But if the electroweak symmetry is intact, the requirement of gauge invariance dictates that composite states, like in QCD, act as the asymptotic states.

This conundrum is resolved by the Fr\"ohlich-Morchio-Strocchi~(FMS) mechanism~\cite{Frohlich:1980gj,Frohlich:1981yi}. This mechanism explains how the Brout-Englert-Higgs~(BEH) effect allows perturbation theory to become the leading term in an expression ordered by the vacuum expectation value. As a consequence, deviations from perturbation theory in the Standard Model are parametrically and kinematically\footnote{This is due to the special group-theoretical structure of the Standard Model and not generally true~\cite{Maas:2017wzi,Maas:2017xzh,Sondenheimer:2019idq}.} suppressed~\cite{Maas:2024hcn,Maas:2017wzi,Dudal:2020uwb,Maas:2020kda}. For the gauge-Higgs sector, this has been confirmed in lattice calculations, see~\cite{Maas:2023emb} for an overview.

For fermions, this scenario has been so far only investigated exploratory~\cite{Aoki:1988fg,Aoki:1988fg,Lee:1987zu,Lee:1988ut,Lee:1987eg} or quenched~\cite{Afferrante:2020fhd,Greensite:2026jim} using non-perturbative lattice methods. Here, we begin to fill this gap. While the underlying FMS mechanism for fermions has been well understood~\cite{Frohlich:1980gj,Frohlich:1981yi,Egger:2017tkd,Afferrante:2020fhd}, effects involving them may be able to test this mechanism at future colliders~\cite{Maas:2022gdb}. Moreover, subleading electroweak physics in future experiments~\cite{deBlas:2025gyz} may be crucially affected. Similar effects have been observed in lattice calculations of the bosonic sector~\cite{Jenny:2022atm,Maas:2018xxu,Greensite:2025fdq}; the present investigation will lay the foundations to do the same in the fermion sector.

That said, a compromise has to be made in practice. As gauged Weyl fermions are (not yet) accessible in lattice calculations~\cite{Hasenfratz:2007dp}, a replication of the lepton (or quark) sector of the Standard Model is not possible on the lattice. Thus, a genuine quantitative prediction remains unfeasible, though the enormous scale differences would not allow it anyhow in practice. Fortunately, the FMS mechanism is transparent to the exact nature of the fermions~\cite{Afferrante:2020fhd}. Thus, qualitative statements like in which observables which order of magnitude effects can be expected remain possible, and will be able to guide analytic calculations -- FMS-augmented perturbation theory~(APT) -- along the lines of~\cite{Dudal:2020uwb,Maas:2020kda,Maas:2024hcn,Maas:2022gdb}.

Here, we will consider the closest approximation to the lepton-weak sector of the Standard Model that is possible to simulate on the lattice: We couple vectorial (Dirac) fermions as stand-ins for leptons to the weak gauge-Higgs sector, consisting of an SU(2)~Yang-Mills theory and a fundamental Higgs doublet. Our aim is to study the ground state spectrum as a function of the fermion mass, which has been investigated using APT in~\cite{Afferrante:2020fhd}. We will describe this system in more detail in section~\ref{sec:system}, alongside the predictions from APT.

For the lattice simulations, we employ the \texttt{HiRep} code~\cite{DelDebbio:2008zf, Hansen:2017mrt, Drach:2025eyg} including scalars and Wilson fermions. We perform a variational analysis to extract the ground state spectrum. The technical details of these simulations are laid out in section~\ref{sec:lattice}. Several technical details are relegated to appendices~\ref{app:pt} to~\ref{app:data}.

Finally, we present the results in section~\ref{sec:results}. We follow here the philosophy to start with infinitely heavy fermions from three starting points well studied\footnote{We note that we have reproduced the starting points using \texttt{HiRep} without dynamical fermions, employing an entirely different code than used in~\cite{Wurtz:2013ova,Maas:2014pba,Jenny:2022atm}, thereby  providing an independent check of their results.} in~\cite{Wurtz:2013ova,Maas:2014pba,Jenny:2022atm}. They differ by strength of the effective gauge coupling and also the lattice spacing, but are all weakly coupled. All starting points have been found to be in the BEH domain. We then lower the fermion mass as far as possible, which is limited by the well-known~\cite{Montvay:1994cy} loss of reflection positivity related to Wilson fermions, see appendix~\ref{app:pt}. However, before this, the fermions always become light enough to open all possible decay channels, and thus sufficiently light to investigate interesting~\cite{Maas:2022gdb,Jenny:2022atm} matrix elements.

We find that all seven investigated channels behave as expected from APT, as discussed in section~\ref{sec:system}. Moreover, we observe that the fermions do not change the dynamics of the system and the bosonic sector is unaltered until additional decay channels with fermions start to open. This is again in line with expectations from APT, but also from perturbation theory: In the BEH domain of the theory, fermions are only affected but do not qualitatively alter the system. The presence of the scalar appears to stabilize the system against the formation of a walking phase, as it is observed in the same theory without scalars~\cite{Andersen:2011yj}.

The findings presented here set the stage for several further results, which will be reported elsewhere. Preliminary ones can be found in~\cite{Wieland:2025tog,Martins:2026eni}.

\section{SU(2)~scalar-fermion-gauge system}
\label{sec:system}

\subsection{The theory}

A detailed description of the theory and background can be found in~\cite{Afferrante:2020fhd,Maas:2017wzi}. For completeness, we will rehearse here only the elements necessary for the present paper. In particular, we will not discuss the Yukawa sector, as we do not simulate it in this work.

We start with an SU(2)~scalar-gauge system with the scalar field in the fundamental representation. The gauge-scalar part of the Lagrangian can be written as
\begin{equation}
\label{eq:lagrangian_gs}
  \mathcal{L}_{GS} = -\frac{1}{4} W_{\mu \nu}^a W^{a, \mu \nu} + \frac{1}{2} \mathrm{tr} \left[ \left( D_\mu X \right)^\dagger \left( D^\mu X \right) \right] - \tilde{\lambda} \left(\frac{1}{2} \mathrm{tr} \left[ X^\dagger X \right] - f^2\right)^2.
\end{equation}
Here, the matrix-valued scalar field~$X$ is used, which is defined as
\begin{equation}
  X = \begin{pmatrix} \bar{\Phi} & \Phi \end{pmatrix} = \begin{pmatrix} \phi_2^* & \phi_1 \\ -\phi_1^* & \phi_2 \end{pmatrix},
\end{equation}
where~$\bar{\Phi}_i = \epsilon_{ij} \phi_j^*$ in terms of the usual vector representation~$\Phi = (\phi_1 \ \ \phi_2)^\top$ of a complex doublet~\cite{Shifman:2012zz}.

The leptons in the Standard Model are added with the Lagrangian
\begin{equation}
  \mathcal{L}_F  =  \sum_g \bar{\psi}_g^L i \slashed{D} \psi_g^L + \sum_f \bar{\chi}_f^R i \slashed{\partial} \chi_f^R .\label{eq:smls}
\end{equation}
Since we include only two generations in the simulations,~$g=1,2$ sums over the left-handed generations of doublets and~$f=1,2,3,4$ over the right-handed flavors. The right-handed leptons decouple entirely without Yukawa interactions (or hypercharge). They will therefore be henceforth ignored. The left-handed doublets can be written as
\begin{equation}
    \psi^L_g = \begin{pmatrix} \nu_l^L & l^L \end{pmatrix}^\top,
\end{equation}
where we label both components according to the conventional choice. Note, however, that in absence of the Yukawa sector, there is no distinction between what is identified as neutrinos and proper leptons aside from a gauge index. In particular, leptons and neutrinos can be transformed into each other locally by a weak gauge transformation, much like red, green, and blue quarks can be transformed into each other.

As noted in the introduction, for the purpose of lattice simulations we need to replace Weyl fermions by Dirac fermions\footnote{However, note~\cite{Gattringer:2008je}.},
\begin{equation}
\label{eq:lagrangian_f}
  \mathcal{L}_F  =  \sum_g \bar{\psi}_g \left(i \slashed{D} - m_{\psi_g} \right) \psi_g + \sum_f \bar{\chi}_f \left( i \slashed{\partial}  - m_{\chi_f} \right) \chi_f.
\end{equation}
The only difference at the Lagrangian level is the possibility of additional Dirac mass terms. This changes neither the weak structure nor the fact that the would-be right-handed fermions are ungauged and decouple, and can therefore still be ignored. Nor does the assignment of lepton and neutrino identities change. In fact, as SU(2) is pseudo-real, particles and anti-particles can be rotated into each other, a fact known as Pauli-G\"ursey symmetry~\cite{vonSmekal:2012vx}. As a consequence, the theory can be rewritten using two times the number of Majorana generations, equivalent to an arbitrary choice of left-handed or right-handed Weyl fermions. Thus, this theory effectively corresponds to the Standard Model lepton sector~\eqref{eq:smls}, but with two left-handed and two right-handed gauged generations\footnote{Note that the introduction of hypercharge will remove the possibility for an ordinary mass term in this theory and requires the Yukawa sector for mass generation.}. In practice, we keep the Dirac formulation, corresponding to an enlarged symmetry which allows not only transformations between different generations at degenerate masses, but also between particles and anti-particles in the same generation~\cite{Kogut:2000ek}.

The total symmetries of the theory are then~$\text{SU(2)}_{\text{gauge}} \times \text{SU(2)}_{\text{Higgs}} \times \text{SU(4)}_{\text{generation}}$, where only SU(2)$_{\text{gauge}}$ is a local symmetry and a global U(1)~symmetry has been removed by the fermion anomaly~\cite{vonSmekal:2012vx}. Gauge-invariant states can be labeled, besides~$J^P$, with the custodial representation of SU(2)$_{\text{Higgs}}$ and the generational representation of SU(4)$_{\text{generation}}$. Degenerate mass terms break the latter symmetry explicitly to Sp(4)$_{\text{generation}}$~\cite{vonSmekal:2012vx}. We will keep this symmetry intact for now, as degenerate masses allow for a better performance of the employed lattice formulation of fermions, avoiding a (mild) sign problem. Thus, generational multiplets need to be arranged in terms of representations of Sp(4). This has been worked out in detail in the case of QCD-like theories~\cite{Kogut:2000ek,vonSmekal:2012vx,Drach:2017btk,Bennett:2019jzz,Kulkarni:2022bvh}.

\subsection{The state space}

As has been discussed in the introduction, any physical states need to be gauge-invariant. In lattice simulations, which do not need to gauge fix, this is doubly true: Gauge-dependent quantities will vanish, if the gauge is not fixed. The possibility of gauge fixing will be addressed in the next section.

While there are an infinite number of possible channels, the FMS mechanism motivates that in the BEH domain of the theory only very few of them feature bound states, or even non-trivial resonances~\cite{Frohlich:1980gj,Frohlich:1981yi,Maas:2017xzh,Egger:2017tkd,Maas:2017wzi}. Without dynamical fermions, lattice results support this expectation~\cite{Wurtz:2013ova,Maas:2014pba,Afferrante:2020fhd,Jenny:2022atm}. We will therefore concentrate on channels in which bound states are expected according to the FMS mechanism, and a few other channels where this is not the case and which are accessible within reasonable computing time. We list these channels, together with sample operators, in table~\ref{tab:channels}. A full list of all operators used in our variational analysis will be given in section~\ref{sec:spec}.

\begin{table}
  \centering
  \begin{tabular}{| c | c || c | c | c |}
  \hline
  Name & $\,$ Sample operator $\,$ & $\, J^P \,$ & $\,$ SU(2)$_{\text{Higgs}}$ $\,$ & $\,$ Sp(4)$_{\text{generation}}$ $\,$ \\
  \hline\hline
    Higgs & $\mathrm{tr} \left[ X^\dagger X \right]$ & $0^+$ & $\mathbf{1}$ & $\mathbf{1}$ \\
    \hline
    W/Z~boson & $\;$$\mathrm{tr} \left[ \tau^{\mathbf{a}} X^\dagger D_\mu X \right]$$\;$  & $1^-$ & $\mathbf{3}$ & $\mathbf{1}$ \\
    \hline
    Fermion & $X^\dagger \psi$  & $\frac{1}{2}^+$ & $\mathbf{2}$ & $\mathbf{2}$ \\
    \hline
    $\,$ Pseudo-scalar $\,$ & $\tau^3_{ij}\bar{\psi}_i\gamma_5\psi_j$ & $0^-$ & $\mathbf{1}$ & $\mathbf{5}$ \\
    \hline
    Scalar & $\tau^3_{ij}\bar{\psi}_i\psi_j$ & $0^+$ & $\mathbf{1}$ & $\mathbf{5}$ \\
    \hline
    Vector & $\tau^3_{ij}\bar{\psi}_i\gamma_\mu\psi_j$ & $1^-$ & $\mathbf{1}$ & $\mathbf{10}$ \\
    \hline
    Axial-vector & $\tau^3_{ij}\bar{\psi}_i\gamma_5\gamma_\mu\psi_j$ & $1^+$ & $\mathbf{1}$ & $\mathbf{10}$ \\
    \hline
  \end{tabular}
  \caption{Physical channels. Note that all of them are gauge singlets.}
  \label{tab:channels}
\end{table}

For the Standard Model case, the fermion channel corresponds to a left-handed lepton, while the bosonic states with non-trivial generation number would be bound states of two leptons of different generations, i.e., leptonia. In the Standard Model, these leptonia are bound by electric interactions, e.g., a muon and a positron in some spin state. In absence of hypercharge, as in the present model, it is expected that these are unbound and that there are only scattering states of twice the fermion mass in these channels with additional necessary relative momenta to generate the spins. This standard expectation of the absence of bound states in all leptonium channels is also true in the FMS picture~\cite{Maas:2017xzh}.

In a standard perturbative setup, all of the (composite) channels in table~\ref{tab:channels} would be scattering states of the constituents. Thus, for the bosonic generation non-singlets the same predictions are made. For the Higgs, W/Z boson, and fermion, however, the prediction is different, giving a mass equal to the sum of the constituents of the scattering states required to make up the quantum numbers. This predicts for the Higgs the mass~$2m_\text{Higgs}$, for the W/Z~boson~$2m_\text{Higgs}+m_\text{W/Z}$, and for the fermion~$m_\text{Higgs} + m_\text{Fermion}$. However, in stark contrast, the FMS mechanism predicts for these states the masses~$m_\text{Higgs}$,~$m_\text{W/Z}$, and~$m_\text{Fermion}$~\cite{Frohlich:1980gj,Frohlich:1981yi,Afferrante:2020fhd,Dudal:2020uwb,Maas:2020kda}. In particular, the FMS mechanism also predicts that the fermion channel can be arbitrarily lighter than the Higgs channel. This is a genuine prediction which will be checked in the following.

The agreement of the masses of the composite states with the elementary masses predicted by the FMS mechanism has been tested for the quenched case, and found to be true~\cite{Maas:2013aia,Afferrante:2020fhd}. However, determining the masses of the elementary states requires gauge fixing and the calculation of the gauge-fixed propagator. The computational costs involved in doing so on our larger lattices and the necessary renormalization and expected mixing make this a non-trivial exercise~\cite{Maas:2013aia,Afferrante:2020fhd}. Thus, we leave this explicit check for future work.

\subsection{Gauge fixing}

There is, however, an issue for which gauge fixing is needed in the following. Nevertheless, it is sufficient to do so in a much more limited setting -- and thus computationally much cheaper -- than would be required for gauge-fixed propagators.

In all existing purely bosonic simulations the mass ordering of the Higgs and W/Z channels has been in one-to-one correspondence with the BEH mode in Landau gauge~\cite{Evertz:1985fc,Maas:2017wzi,Afferrante:2020hqe,Maas:2018xxu}, whenever this was investigated. Even in the region of the phase diagram, where the onset of the BEH effect differs in different gauges~\cite{Caudy:2007sf}, available results indicate a degeneracy in the Higgs and W/Z channels~\cite{Maas:2014pba}. But it is not a~priori clear whether this remains true in the presence of dynamical fermions. In addition, the origin of this remarkable result is entirely unknown~\cite{Maas:2017wzi}, despite the fact that is has tremendous impact, e.g., on the structure of the thermodynamic phase diagram~\cite{Kapusta:2006pm} and thus baryogenesis~\cite{Bambi:2015mba}.

To monitor the status of the BEH effect, we therefore determine the vacuum expectation value~$v$ of the elementary Higgs field~$\phi$ (or~$X$) in minimal 't~Hooft-Landau gauge~\cite{Maas:2013aia,Maas:2012tj,Maas:2017wzi}. To this end, we fix the gauge in our lattice simulations as described below, and then measure the space-time averaged Higgs field. This object is also rotated per configuration to a fixed direction, so as to be equivalent to the usual $R_\xi$~gauges of perturbation theory~\cite{Bohm:2001yx,Maas:2013aia,Maas:2012ct}. Indeed, we find that the vacuum expectation value is non-zero in the infinite volume limit and basically independent of the fermion mass. This ensures that we stay in the BEH domain throughout our simulations.

This also allows us to define a renormalization scheme in terms of the measured vacuum expectation value~$v$ for the parameters of our theory, which coincides at tree-level with the usual definitions~\cite{Bohm:2001yx},
\begin{align}
    \label{eq:g}
    g&=\frac{m_\text{W/Z}}{v},\\
    \label{eq:lam}
    \tilde{\lambda}&=\left(\frac{m_\text{Higgs}}{v}\right)^2,\\
    m_\psi&=m_\text{Fermion},
\end{align}
yielding the tree-level gauge coupling, the four-Higgs coupling, and the fermion mass in terms of the measured masses in the different channels. We note that the fourth parameter is used to set our scale, by specifying for physical units that always $m_\text{W/Z}=80.375$~GeV. While we do reach fermion masses for which the W/Z~boson becomes unstable against the decay into two fermions, the constancy of the lowest energy level in this channel suggests that the resonance mass remains close to threshold, allowing the continued use of this scale. Of course, a full resonance analysis is needed to confirm this view, which is left for future investigations.

For later comparison, it is useful to note that for the Standard Model parameters (choosing~$m_{W/Z}$ to be the measured W~mass),~$g_\text{SM}=0.32$ (corresponding to~$\alpha_\text{SM}=g^2/(4\pi)=0.0081$) and~$\tilde{\lambda}_\text{SM}=0.25$.

\section{Lattice methods}
\label{sec:lattice}

We simulate the theory defined in equations~\eqref{eq:lagrangian_gs} and~\eqref{eq:lagrangian_f} using the \texttt{HiRep} code base~\cite{DelDebbio:2008zf, Hansen:2017mrt, Drach:2025eyg}. Since we set Yukawa couplings to zero the ungauged fermions decouple and we are left with gauged fermions only. The full lattice action can be decomposed into three parts:
\begin{equation}
    S=S_{\textrm{gauge}}+S_{\textrm{fermion}}+S_{\textrm{scalar}}.
\end{equation}
The pure gauge part is given by
\begin{equation}
    S_{\textrm{gauge}} = \beta \sum_x\left[ 1 - \frac{1}{2}\sum_{\mu < \nu} \textrm{Re} \, \mathrm{tr} \, P_{\mu\nu}(x) \right],
\end{equation}
where~$P_{\mu\nu}(x)$ is the standard elementary $1\times1$~plaquette and~$\beta$ is the bare gauge coupling. Adding two mass-degenerate generations of Wilson fermions with bare mass parameter~$am_F$ requires
\begin{equation}
    S_{\textrm{fermion}} = \sum_{g=1}^2\sum_x \bar{\psi}_g(x) \Big[ \left( 4 + am_F \right) \psi_g(x) - \frac{1}{2} \sum_\mu \big( (1-\gamma^\mu) U_\mu(x)\psi_g(x+\hat{\mu}) + (1+\gamma^\mu)  U^\dagger_\mu(x-\hat{\mu})\psi_g(x-\hat{\mu})  \big) \Big].
\end{equation}
Finally, scalar fields are implemented via
\begin{equation}
    S_{\textrm{scalar}} = \sum_x \Big[ \left( (am_S)^2+8 \right) \phi^\dagger(x) \phi(x) + \lambda \left( \phi^\dagger(x) \phi(x) \right)^2 - \sum_\mu\left( \phi^\dagger(x) U_\mu(x) \phi(x + \hat{\mu}) + \phi^\dagger(x) U^\dagger_\mu(x - \hat{\mu}) \phi(x - \hat{\mu}) \right) \Big]
\end{equation}
with the bare lattice parameters~$(am_S)^2$ and~$\lambda$.
Here, we have switched to the standard lattice versions of the gauge, fermion, and scalar fields ($U_\mu$,~$\psi$, and~$\phi$~/~$X$). This notation will be kept for the rest of this section.

Besides the spectroscopic observables that will be discussed in section~\ref{sec:spec}, we also measure the average plaquette~$\langle P \rangle$, the average scalar length~$\langle \sqrt{\phi^\dagger \phi} \rangle$, and the vacuum expectation value 
\begin{equation}
    av=\bigg\langle\bigg\vert \frac{1}{V} \sum_x \phi(x) \bigg\vert\bigg\rangle,
\end{equation}
where the sum is over all lattice sites. To determine~$av$, we fix to minimal Landau gauge~\cite{Cucchieri:1995pn} using the Cabibbo-Marinari method~\cite{Cabibbo:1982zn}. To fix the gauge completely, the direction of~$\sum_x \phi(x)$ has to be aligned across different configurations, resulting in minimal 't~Hooft-Landau gauge. Values for~$\langle P \rangle$,~$\langle \sqrt{\phi^\dagger \phi} \rangle$, and~$av$ for all ensembles can be found in appendix~\ref{app:data}.

We also measure the pseudo-scalar decay constant. However, as this quantity is not directly relevant to the discussion of our findings, details on its calculation and results are relegated to appendix~\ref{app:fps}. Regardless, its behavior aligns with expectations in the BEH domain.

\subsection{Spectroscopic observables}
\label{sec:spec}

For the spectroscopy, we define lattice versions of the sample operators depicted in table~\ref{tab:channels}. The operators for the scalar, vector, and fermion channels are given by
\begin{align}
\label{eq:ops1}
  \mathcal{O}_{0^+}(x) & = \frac{1}{2} \mathrm{tr} \left[ X^\dagger(x) X(x) \right] \\
\label{eq:ops2} 
  \mathcal{O}_{1^-}^{\mathbf{a}, \mu}(x) & = -\frac{i}{2} \mathrm{tr} \left[ \tau^{\mathbf{a}}X^\dagger(x) U^\mu(x) X(x+\hat{\mu}) \right], \\
\label{eq:ops3}
  \mathcal{O}^{\mathbf{i}, \alpha}_{\Psi}(x) & = X^\dagger_{\mathbf{i} i}(x) \psi^{i, \alpha}(x),
\end{align}
where~$\mathbf{a}$ and~$\mathbf{i}$ are indices belonging to the global SU(2)$_{\text{Higgs}}$ group, while~$\alpha$ encodes the Dirac structure.~$\tau^{\mathbf{a}}$ are the Pauli matrices representing the generators of SU(2)$_{\text{Higgs}}$. These operators are per construction gauge-invariant and can therefore be measured on the lattice without gauge fixing.

To extract masses in the leptonium channels, we use a standard meson operator known from lattice QCD~\cite{Gattringer:2010zz},
\begin{equation}
  \label{eq:lept}
  \mathcal{O}_\Gamma(x) = \bar{\psi}_1(x) \Gamma \psi_1(x) - \bar{\psi}_2(x) \Gamma \psi_2(x).
\end{equation}
The subscripts are generation indices. We consider four channels classified by~$\Gamma$, see table~\ref{tab:channels}; note that different~$\gamma_\mu$ are summed over. After performing all contractions when constructing the corresponding correlators with degenerate generations, only diagrams of the form~$\mathrm{tr} \left[ \Gamma D^{-1} \bar{\Gamma} D^{-1} \right]$ with $\bar{\Gamma} = \gamma_0 \Gamma^\dagger \gamma_0$ survive. Other operators such as~$\bar{\psi}_1 \Gamma \psi_2$ can be used as an alternative, since for degenerate generations the resulting diagrams are identical.

For the present analysis, we are only interested in the ground states of all considered channels. Therefore, we adopt two strategies to remove excited state contaminations. First, we apply APE smearing~\cite{APE:1987ehd,Philipsen:1996af,Maas:2014pba} to all bosonic fields before constructing the operators defined in equations~\eqref{eq:ops1} to~\eqref{eq:ops3}. We measure the operators after~$N_{\text{APE}}=4$ smearing steps with~$\alpha_{\text{APE}} = 0.45$. Second, we perform a variational analysis~\cite{Gattringer:2010zz} in all channels by solving an eigenvalue problem~(EVP) of the form 
\begin{equation}
    \mathbf{C}_{ij}(t) v(t) = \lambda(t) v(t).
\end{equation}
The cross-correlator matrices~$\mathbf{C}_{ij}$ are constructed from the operator bases defined in table~\ref{tab:evp} for momenta~${\bf p=0}$. For every channel, this results in a $3\times 3$~matrix for every Euclidean time~$t$. To avoid level crossings, we sort the eigenvalues according to the overlap~$v_i^\dagger(t) v_j(0)$ for all~$t$. The resulting improved correlation functions~$\lambda(t)$ are fitted according to the procedures described in section~\ref{sec:methods}. 

To verify the resulting ground state energy levels, we compare parts of our EVP results with those obtained from a generalized eigenvalue problem~(GEVP) using extended bases for the scalar, vector, and fermion channels (see appendix~\ref{app:gevp}). The cross-correlator matrices for all channels are available in the dedicated data release~\cite{data_release}.

\begin{table}
  \centering
  \begin{tabular}{| c || c | c | c | c |} 
  \hline
    Channel & $\,$ Scalar singlet $\,$ & $\,$ Vector triplet $\,$ & $\,$ Fermion doublet $\,$ & $\,$ Leptonia $\,$ \\
    \hline\hline
    \multirow{3}{6.9em}{\centering Basis elements} & $\mathcal{O}_{0^+}$ & $\mathcal{O}^{{\bf a}, \mu}_{1^-}$ & $\mathcal{O}^{\mathbf{i}, \alpha}_{\Psi}$ & $\mathcal{O}_{\Gamma}$ \\
    & $\mathcal{O}_{0^+} \, \mathcal{O}_{0^+}$ & $\mathcal{O}_{0^+} \, \mathcal{O}^{{\bf a}, \mu}_{1^-}$ & $\mathcal{O}_{0^+} \, \mathcal{O}^{\mathbf{i}, \alpha}_{\Psi}$ & $\mathcal{O}_{0^+} \, \mathcal{O}_{\Gamma}$ \\
    & $\mathcal{O}^{{\bf b}, \nu}_{1^-} \, \mathcal{O}^{{\bf b}, \nu}_{1^-}$ & $\mathcal{O}^{{\bf b}, \nu}_{1^-} \mathcal{O}^{{\bf b}, \nu}_{1^-}  \, \mathcal{O}^{{\bf a}, \mu}_{1^-}$ & $\mathcal{O}^{{\bf b}, \nu}_{1^-} \mathcal{O}^{{\bf b}, \nu}_{1^-} \, \mathcal{O}^{\mathbf{i}, \alpha}_{\Psi}$ & $\mathcal{O}^{{\bf b}, \nu}_{1^-} \mathcal{O}^{{\bf b}, \nu}_{1^-} \, \mathcal{O}_{\Gamma}$ \\[2pt]
    \hline
  \end{tabular}
  \caption{EVP operator bases for all channels of interest.}
  \label{tab:evp}
\end{table}

\subsection{Spectroscopy methods}
\label{sec:methods}

To obtain states of definite spatial momentum~${\bf p}$, we need to perform a Fourier transformation. Since we are only interested in~${\bf p=0}$ operators, this simplifies to a sum over spatial lattice sites,
\begin{equation}
 \mathcal{O}(t) = \frac{1}{V} \sum_{\mathbf{x}} \mathcal{O}(t,\mathbf{x}).
\end{equation}
The operators in the scalar singlet channel (second column in table~\ref{tab:evp}) mix with the vacuum. Therefore, we subtract the disconnected contributions by replacing~$\mathcal{O}_{0^+}(t)$ with~$\mathcal{O}_{0^+}(t) - \langle \mathcal{O}_{0^+}(t) \rangle$. To reduce the statistical noise in the bosonic channels, we use time-slice averaging when constructing purely bosonic cross-correlators,
\begin{equation}
  \mathbf{C}^{\text{B}}_{ij}(t) = \frac{1}{L_t} \sum_{\tau=0}^{L_t-1} \langle \mathcal{O}_i (\tau + t) \mathcal{O}_j^\dagger (\tau) \rangle.
\end{equation}
$L_t$ is the temporal lattice extent.

For operators including fermion fields, standard correlators of the form
\begin{equation}
  \mathbf{C}^{\text{F}}_{ij}(t) = \langle \mathcal{O}_i(t) \bar{\mathcal{O}}_j(0) \rangle
\end{equation}
are constructed. This allows us to use point sources for inverting the Dirac propagator. Due to the choice of bases, the cross-correlator matrices in both the fermion and leptonium channels factorize. For the fermion doublet, this results in the following matrix,
\begin{equation}
  \mathbf{C}_{\Psi}^{\text{F}}(t) =
  \begin{pmatrix}
    f_\Psi(t \vert 0) & f_\Psi(t \vert 0) f^*_{0^+}(0) & f_\Psi(t \vert 0) f^*_{1^-}(0) \\
    f_{0^+}(t) f_\Psi(t \vert 0) & f_{0^+}(t) f_\Psi(t \vert 0) f^*_{0^+}(0) & f_{0^+}(t) f_\Psi(t \vert 0) f^*_{1^-}(0)  \\
    f_{1^-}(t) f_\Psi(t \vert 0)  & f_{1^-}(t) f_\Psi(t \vert 0) f^*_{0^+}(0) & f_{1^-}(t) f_\Psi(t \vert 0) f^*_{1^-}(0)  \\
  \end{pmatrix}.
\end{equation}
Analogously, the leptonium EVP factorizes as 
\begin{equation}
  \mathbf{C}^{\text{F}}_{\Gamma}(t) =
  \begin{pmatrix}
    f_\Gamma(t \vert 0) & f_\Gamma(t \vert 0) f^*_{0^+}(0) & f_\Gamma(t \vert 0) f^*_{1^-}(0) \\
    f_{0^+}(t) f_\Gamma(t \vert 0) & f_{0^+}(t) f_\Gamma(t \vert 0) f^*_{0^+}(0) & f_{0^+}(t) f_\Gamma(t \vert 0) f^*_{1^-}(0) \\
    f_{1^-}(t) f_\Gamma(t \vert 0) & f_{1^-}(t) f_\Gamma(t \vert 0) f^*_{0^+}(0) & f_{1^-}(t) f_\Gamma(t \vert 0) f^*_{1^-}(0) \\
  \end{pmatrix}
\end{equation}
for all~$\Gamma$. Due to this factorization, only the following four expressions have to be measured to construct the fermionic cross-correlator matrices,
\begin{align}
  f_\Psi(t \vert 0) & = \frac{1}{2V} \sum_{\mathbf{x}}\mathrm{tr} \left[ X^\dagger(t,\mathbf{x}) D^{-1}(t,\mathbf{x} \vert 0,0) X(0,0) \right] ,\\
  f_\Gamma(t\vert 0) & = -\frac{1}{V} \sum_{\mathbf{x}} \mathrm{tr} \left[ \gamma_5 \Gamma D^{-1}(t, \mathbf{x} \vert 0,0) \bar{\Gamma} \gamma_5 D^{-1}(t, \mathbf{x} \vert 0,0) \right] ,\\
  f_{0^+}(t) & = \frac{1}{V} \sum_{\mathbf{x}} \mathcal{O}_{0^+}(t, \mathbf{x}), \\
  f_{1^-}(t) & = \frac{1}{V^2} \sum_{\mathbf{x}, \mathbf{y}} \sum_{\mathbf{b}, \nu} \mathcal{O}_{1^-}^{\mathbf{b}, \nu}(t, \mathbf{x}) \mathcal{O}_{1^-}^{\mathbf{b}, \nu}(t,\mathbf{y}).
\end{align}
Note that the Dirac operator~$D^{-1}(t,\mathbf{x} \vert 0,0)$ has to be inverted only once.

To reduce autocorrelations, we skip~$L_t/2$ configurations between measurements and add~$N_{\textrm{bin}}$ binning steps after thermalization. The number of binning steps is estimated via~$N_{\textrm{bin}} \geq 2\tau_{\textrm{ac}}$, where~$\tau_{\textrm{ac}}$ is the autocorrelation time calculated using the Julia package \texttt{ADerrors}~\cite{Ramos:2018vgu}. Exact values for~$N_{\textrm{bin}}$ can be found in appendix~\ref{app:data}.

For solving the EVP and fitting the resulting correlators, error propagation is performed using statistical bootstrap~\cite{Gattringer:2010zz}. To this end, $\mathcal{O}(1000)$~bootstrap samples are created. For every sample, the seven EVPs are solved with arbitrary precision and the resulting correlators are fitted using standard linear least squares with single or double cosh functions given by
\begin{align}
    C_{\textrm{single}}(t) & = A \cosh \left[m \left( t - \frac{L_t}{2} \right)\right] \\
    C_{\textrm{double}}(t) & = A \, \bigg\{ \cosh \left[m \left( t - \frac{L_t}{2} \right)\right] + B^2 \cosh \left[\left( \vert m \vert + \Delta m \right) \left( t - \frac{L_t}{2} \right)\right] \bigg\}.
\end{align}
For the fits in the scalar and axial-vector leptonium channels, constants have to be added to the fitting functions to account for intersections with the $t$-axis. For very massive leptonium correlators, weighted fits proved to be useful.

To determine the ideal fitting function, both single and double cosh fits are performed and compared. The best fit is then chosen automatically by checking the number of available data points, the goodness of both fits, and the overlap between first and second energy level of the double cosh. This ensures that single cosh fits are allowed especially when almost all information on excited states is lost due to the smearing procedure or the variational analysis. The fitting range~$\left[t_{\textrm{min}}, t_{\textrm{max}}\right]$ is also determined automatically. The lower bound~$t_{\textrm{min}}$ is chosen to be as small as possible while still allowing for an accurate fit of the center of the lattice. The upper bound~$t_{\textrm{max}}$ is chosen to be maximal as long as the correlator is monotonic and the error of all included points is below 50~\%.

To verify the ground state masses of the FMS operators, we also implement a second fitting procedure described in~\cite{Jenny:2022atm}, which provides an automated cross-check for our single and double cosh fits.

Finally, to determine the infinite volume masses we fit the ground state energy levels for finite lattice extent~$L$ with the ansatz~\cite{Luscher:1985dn}
\begin{equation}
\label{eq:iv}
    m(L) = m_\infty + \frac{a}{L} e^{-bL}.
\end{equation}
For these fits, we use a Bayesian inference based method using the Julia package \texttt{Turing}~\cite{10.1145/3711897,pmlr-v84-ge18b}. For this procedure,~$m(L)$ and~$m_\infty$ are assumed to follow a Gaussian distribution. Details on the infinite volume behavior and example plots for the infinite volume fits are presented in appendix~\ref{app:ivfits}.

\subsection{Parameter sets}
\label{sec:sets}

As already mentioned in the introduction, the strategy for simulating the SU(2)~scalar-fermion-gauge system is to choose different starting points already investigated for the gauge-scalar subsystem. Therefore, we have chosen three parameter sets in the BEH domain from lattice calculations published in~\cite{Jenny:2022atm,Wurtz:2013ova}, which differ by gauge coupling and lattice spacing. One set (set~2) has an unstable Higgs, which has been identified by a L\"uscher analysis~\cite{Jenny:2022atm}; the other two (sets~1 and~3) have a stable Higgs, but with different masses. Set~3 from~\cite{Wurtz:2013ova} is the one at weakest coupling and is most similar to the Standard Model. The parameters for the gauge-scalar system compatible with \texttt{HiRep} input requirements are provided in table~\ref{tab:sets}.

\begin{table}
  \centering
  \begin{tabular}{| l | c || c | c | c || c | c | c || c |}
    \hline
    $\,$ Name $\,$ & $\,$ Set $\,$ & $\beta$ & $(am_S)^2$  & $\lambda$ & $g$ & $ \alpha_W $ & $\tilde{\lambda}$ & $\,$ Source $\,$ \\
    \hline\hline
    $\,$ Stable Higgs & s1 & $\, 2.8859 \,$ & $\, -13.5954 \,$ & $\, 15.0118 \,$ & $\, 1.18 \,$ & $\, 0.111 \,$ & $\, 5.04 \,$ &~\cite{Jenny:2022atm} \\
    \hline
    $\,$ Higgs-like resonance $\,$ & s2 & $\,4.0000\,$ & $\,-11.3333\,$ & $\,11.1111\,$ & $\, 0.87 \,$ & $\, 0.061 \,$ & $\, 8.90 \,$ &~\cite{Jenny:2022atm} \\
    \hline
    $\,$ Weak coupling & s3 & $\,8.0000\,$ & $\,-0.4168\,$ & $\,0.1923\,$ & $\, 0.54 \,$ & $\, 0.023 \,$ & $\, 0.69 \,$ &~\cite{Wurtz:2013ova} \\
    \hline
  \end{tabular}
  \caption{Gauge-scalar parameter sets with lattice parameters~$\beta$,~$(am_S)^2$, and~$\lambda$. Symmetric lattice volumes~$L^4$ with~$L=12,16,20,24,28$ are available for all sets. For set~3, additional simulations with~$L=32$ are performed. Estimated values for the tree-level gauge coupling~$g$ (or~$\alpha_W$), defined in equation~\eqref{eq:g}, and the four-Higgs coupling~$\tilde{\lambda}$, defined in equation~\eqref{eq:lam}, are also provided. These values are determined by averaging the vacuum expectation value and the Higgs mass over all available fermion masses and should allow for a comparison with the Standard Model parameters. For set~2, the resonant Higgs mass was taken from~\cite{Jenny:2022atm}.}
  \label{tab:sets}
\end{table}

Starting from simulations of the gauge-scalar theory, dynamical Wilson fermions are added by reducing the bare fermion mass parameter step-by-step. From a first run of exploratory calculations, we were able to determine five simulation points per set in terms of the bare fermion mass parameter. These points will allow us to explore different decay channels in a future scattering analysis. The corresponding values for the bare fermion mass can be found in table~\ref{tab:params}.

\begin{table}
  \centering
  \begin{tabular}{| l | c || c | c | c | c | c |}
    \hline
    $\,$ Name & $\,$ Set $\,$ & sp1 & sp2 & sp3 & sp4 & sp5 \\
    \hline\hline
    $\,$ Stable Higgs & s1 & $\,0.5045\,$ & $\,-0.02544\,$ & $\,-0.1479\,$ & $\,-0.2491\,$ & $\,-0.3768\,$ \\
    \hline
    $\,$ Higgs-like resonance $\,$ & s2 & $\,2.4020\,$ & $\,0.6296\,$ & $\,0.3029\,$ & $\,0.09836\,$ & $\,-0.1774\,$ \\
    \hline
    $\,$ Weak coupling & s3 & $\,0.3103\,$ & $\,0.06174\,$ & $\,-0.009577\,$ & $\,-0.02544\,$ & $\,-0.06299\,$ \\
    \hline
  \end{tabular}
  \caption{Bare fermion masses~$am_F$ for parameter sets~1,~2, and~3.}
  \label{tab:params}
\end{table}

\section{Results}
\label{sec:results}

\subsection{Exploring the parameter space}
\label{sec:expl}

Before discussing the spectroscopy results for the FMS observables and the different leptonium channels, it is worthwhile to dedicate this subsection to a more general discussion on the exploration of the parameter space.

\begin{figure}
    \centering
    \includegraphics[width=0.49\linewidth]{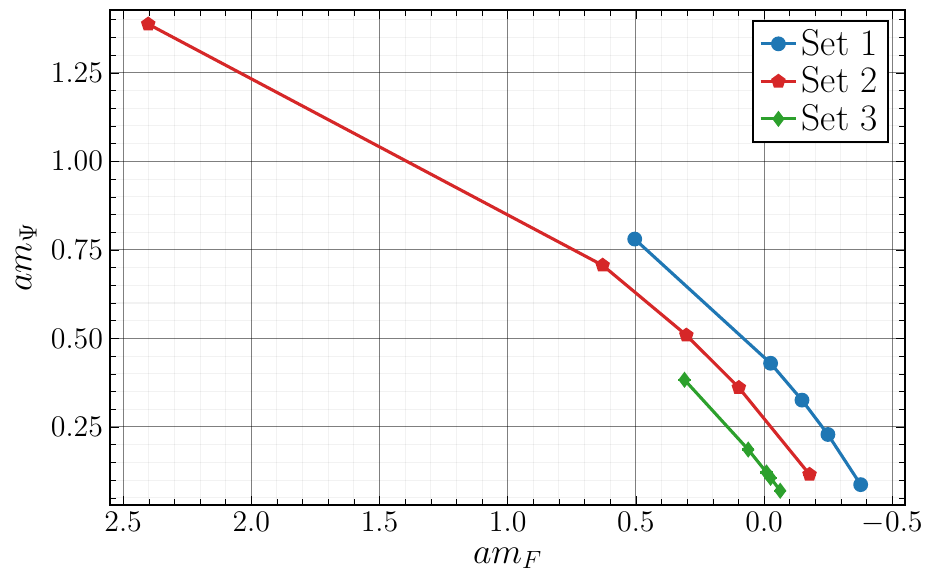} \hfill
    \includegraphics[width=0.49\linewidth]{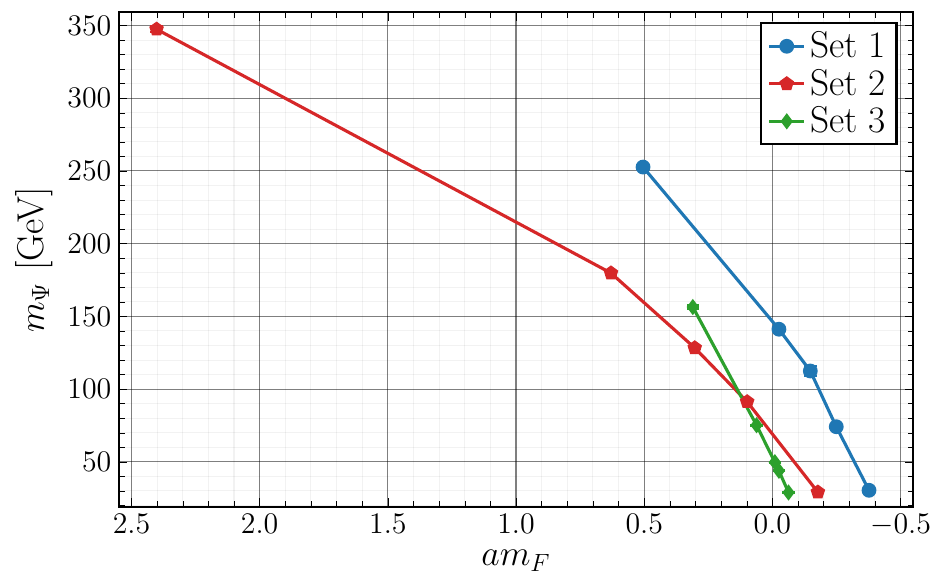}
    \caption{Fermion mass in lattice units (left) and physical units (right) as a function of the bare fermion mass parameter~$am_F$ for the three parameter sets~1,~2, and~3. The plots show an identical behavior for all three sets that allows for a bijective map between~$am_F$ and~$m_\Psi$.}
    \label{fig:mF}
\end{figure}

The first point in this discussion is best illustrated with figure~\ref{fig:mF}. For all sets, the fermion mass is in a one-to-one relation with the corresponding lattice parameter. This is not only true in lattice units but also after setting the scale via the W/Z~boson mass. This is the first hint for a stable gauge-scalar subsystem, which is almost independent of the fermion mass. Furthermore, this bijective relation between~$am_F$ and the gauge-invariant fermion mass~$m_\Psi$ allows us to represent all results in terms of~$m_\Psi$. This is especially useful since the fermion and leptonium masses are the only quantities that show significant changes over different ensembles (simulation points) of the individual sets.

\begin{figure}
    \centering
    \includegraphics[width=0.49\linewidth]{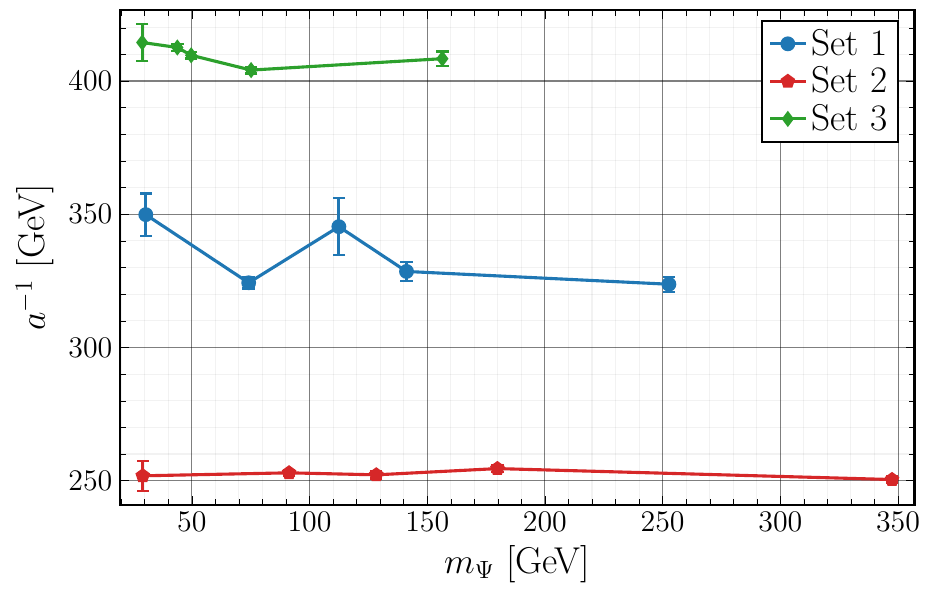} \hfill
    \includegraphics[width=0.49\linewidth]{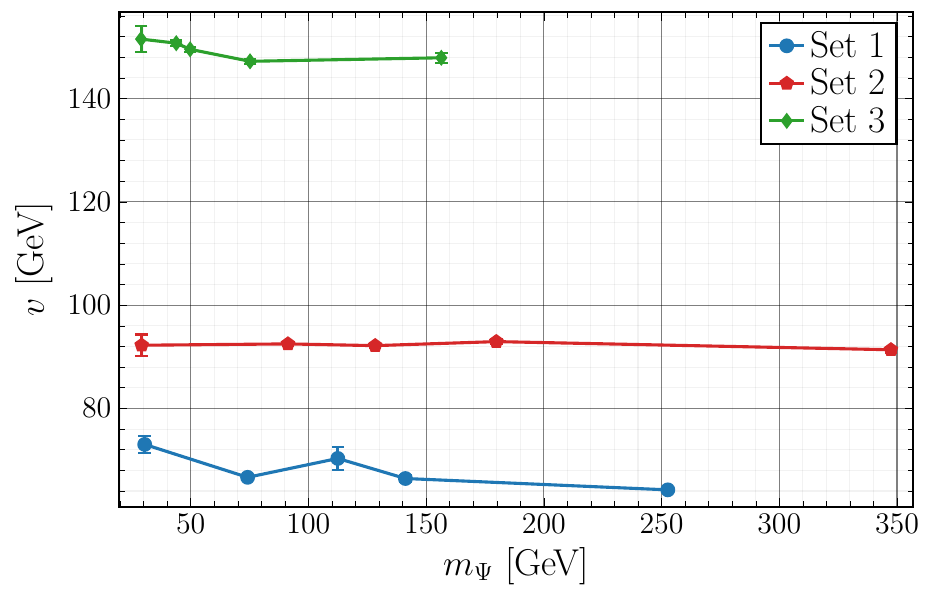} 
    \caption{Lattice spacing~$a^{-1}$ (left) and gauge-fixed vacuum expectation value~$v$ (right) as a function of the fermion mass for all parameter sets. The infinite volume extrapolation for the vacuum expectation value is performed via~$v + c/L^3$. For all sets, this infinite volume fit is dominated by a constant contribution.}
    \label{fig:vev}
\end{figure}

To investigate the stability of the gauge-scalar dynamics further, we first want to focus on changes in the lattice spacing~$a$. The left plot in figure~\ref{fig:vev} depicts the dependence of~$a^{-1}$ on the fermion mass. For all sets, the lattice spacing is approximately constant within errors. Note that due to our scale setting method~$a^{-1}$ is inversely proportional to the vector triplet mass~$m_{1^-}$. The right plot in figure~\ref{fig:vev} shows the vacuum expectation value in physical units which exhibits a similar constant behavior. This not only supports the statement on the stability of the dynamics of the gauge-scalar subsystem but also verifies that we are in the BEH domain of the theory.

The infinite volume extrapolated ground state masses as well as the lattice spacings and the vacuum expectation values for all ensembles can be found in table~\ref{tab:data_iv}. The ground state energy levels of the FMS observables for finite lattice extent can be found in tables~\ref{tab:data_s1} to~\ref{tab:data_s3} in the appendix. The finite lattice results for the leptonium ground state energy levels are not provided in this paper but can be obtained through the dedicated data release~\cite{data_release}.

\begin{table}
\centering
\begin{tabular}{| c || c | c | c | c || c | c | c | c |}
    \hline
    Ensemble & $a^{-1}$ & $v$ & $m_{0^+}$ & $m_{\Psi}$ & $m_{\mathrm{PS}}$ &   $m_{\mathrm{S}}$ & $m_{\mathrm{V}}$ & $m_{\mathrm{AV}}$ \\
    \hline\hline
    s1-sp1 & $324(3)$ & $64.2(6)$ & $159(7)$ & $253(3)$ & $514(12)$ & $564(7)$ & $499(6)$ & $550(20)$ \\
    \hline
    s1-sp2 & $329(4)$ & $66.4(8)$ & $149(7)$ & $141(2)$ & $291(9)$ & $318(9)$ & $301(4)$ & $290(30)$ \\
    \hline
    s1-sp3 & $346(12)$ & $70(3)$ & $155(9)$ & $113(4)$ & $253(9)$ & $266(11)$ & $265(9)$ & $278(11)$ \\
    \hline
    s1-sp4 & $324(3)$ & $66.7(5)$ & $155(6)$ & $74.2(6)$ & $168.6(1.2)$ & $180(9)$ & $172.0(1.3)$ & $177(10)$ \\
    \hline
    s1-sp5 & $350(8)$ & $73(2)$ & $146(10)$ & $30.5(7)$ & $65(2)$ & $111(12)$ & $63(2)$ & $97(10)$ \\
    \hline\hline
    s2-sp1 & $250.4(1.4)$ & $91.4(6)$ & $170(4)$ & $347(2)$ & $715(7)$ & $712(11)$ & $726(5)$ & $750(30)$ \\
    \hline
    s2-sp2 & $254.5(1.4)$ & $92.9(5)$ & $167(3)$ & $179.8(1.0)$ & $347(4)$ & $380(20)$ & $361(3)$ & $380(20)$ \\
    \hline
    s2-sp3 & $252(2)$ & $92.1(6)$ & $164(4)$ & $128.4(8)$ & $248(5)$ & $238(14)$ & $289(4)$ & $235(14)$ \\
    \hline
    s2-sp4 & $252.9(1.0)$ & $92.5(4)$ & $169(4)$ & $91.3(4)$ & $197(2)$ & $214(4)$ & $219.1(9)$ & $209(4)$ \\
    \hline
    s2-sp5 & $252(6)$ & $92(3)$ & $164(6)$ & $29.2(7)$ & $63(2)$ & $72(7)$ & $64(2)$ & $66(6)$ \\
    \hline\hline
    s3-sp1 & $408(3)$ & $147.9(1.0)$ & $123(3)$ & $156.5(1.1)$ & $337(3)$ & $378(3)$ & $344(3)$ & $363(3)$ \\
    \hline
    s3-sp2 & $404.1(1.3)$ & $147.2(5)$ & $124.8(1.1)$ & $75.2(3)$ & $166.1(7)$ & $201(2)$ & $173.4(6)$ & $171(5)$ \\
    \hline
    s3-sp3 & $409.7(1.3)$ & $149.5(5)$ & $123.1(1.0)$ & $49.8(2)$ & $112.5(4)$ & $116(7)$ & $111.9(4)$ & $158(3)$ \\
    \hline
    s3-sp4 & $412.6(1.4)$ & $150.7(6)$ & $126.9(1.3)$ & $43.9(2)$ & $98.6(4)$ & $86(6)$ & $97.7(4)$ & $88(5)$ \\
    \hline
    s3-sp5 & $415(7)$ & $152(3)$ & $124(3)$ & $29.0(5)$ & $66.9(1.1)$ & $71(5)$ & $62.3(1.1)$ & $73(5)$ \\
    \hline
\end{tabular}
\caption{Lattice spacing, vacuum expectation value, and infinite volume extrapolated ground state masses for parameter sets~1,~2, and~3. All quantities are given in GeV. Only statistical uncertainties are shown.}
\label{tab:data_iv}
\end{table}

\subsection{Energy spectra of the FMS observables}
\label{sec:elvs}

The primary objective of this paper is to determine the ground state energy spectra of the FMS observables, i.e., the Higgs, W/Z~boson, and lepton equivalents of the theory discussed in section~\ref{sec:system}. For all three sets, we are able to extract the ground state masses in the scalar singlet, vector triplet, and fermion doublet channels using smearing techniques and a variational analysis. We find compatible results when changing the operator bases and performing a GEVP instead of an EVP analysis (see appendix~\ref{app:gevp}), which further substantiates this claim.

By comparing two different variational analyses and using two different fitting routines for determining the EVP energy levels at finite lattice extent, we are able to exclude most systematic uncertainties related to the extraction of the energy levels. By furthermore carefully performing the infinite volume fits, see appendix~\ref{app:ivfits}, we assume that the main contribution to the errors of the masses in the FMS channels are statistical uncertainties\footnote{In addition, there are systematic effects due to the lattice spacing, which is not varied here. This will require a determination of lines of constant physics, an orders of magnitude more expensive operation in terms of computing power, and is thus beyond our current scope. However, investigations in the bosonic sector suggest that these effects will likely be small or negligible when moving to smaller~$a$~\cite{Maas:2014pba}.}.

\begin{figure}
    \centering
    \includegraphics[width=0.49\linewidth]{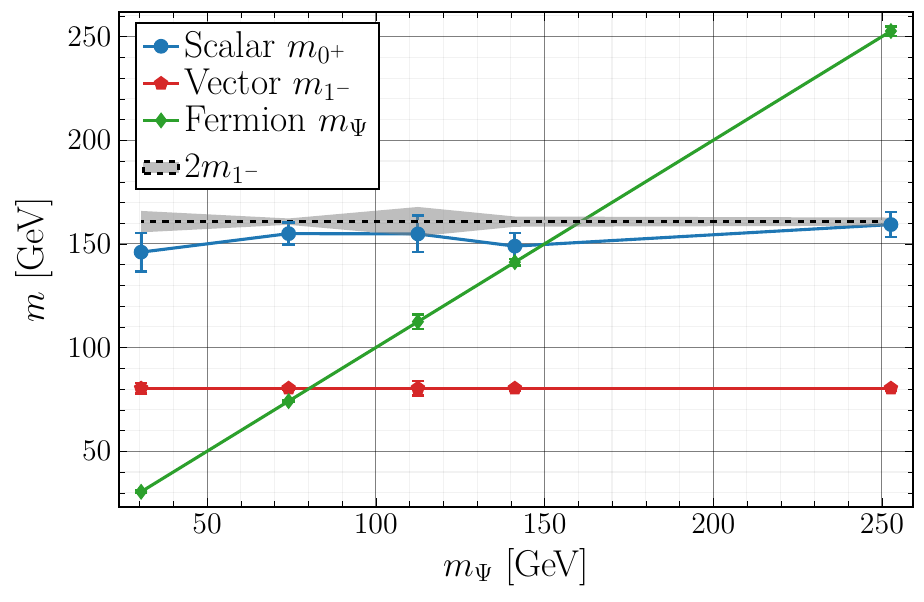} \hfill
    \includegraphics[width=0.49\linewidth]{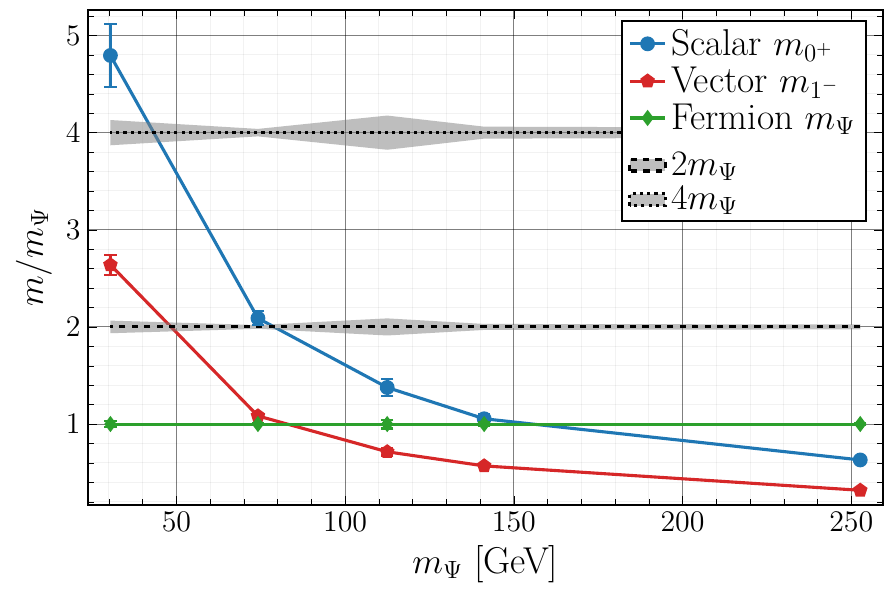}
    \caption{Physical spectrum for the stable Higgs set (set~1). The left plot depicts the masses as a result of the EVP procedure and a subsequent infinite volume fit against the physical lepton mass~$m_\Psi$. The plot on the right shows the ratio~$m/m_\Psi$ including the~$2m_\Psi$ and~$4m_\Psi$ thresholds. Due to the choice of lattice parameters, the $0^+$~mass is close to but still compatible with being below the~$2m_{1^-}$ threshold.}
    \label{fig:s1}
\end{figure}

\begin{figure}
    \centering
    \includegraphics[width=0.49\linewidth]{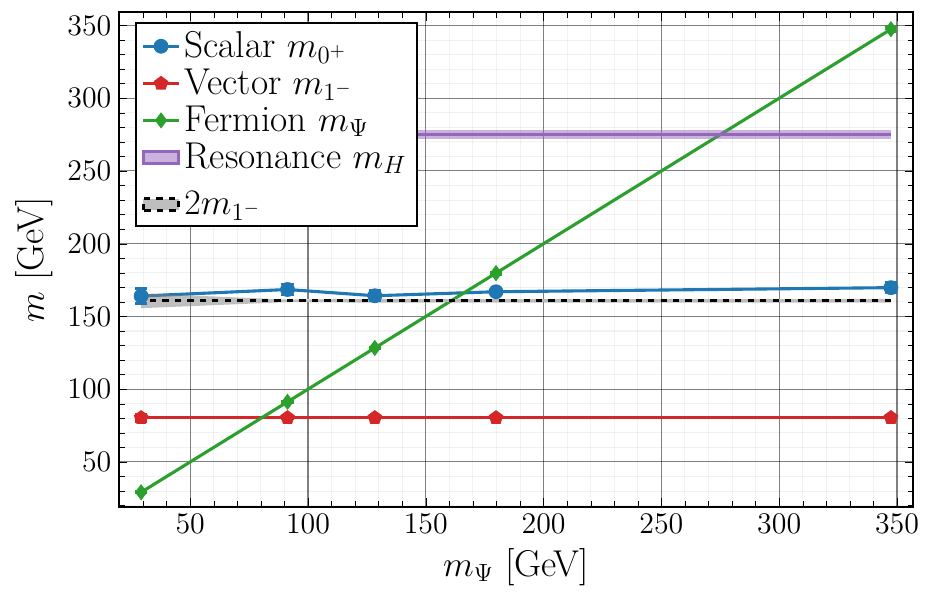} \hfill
    \includegraphics[width=0.49\linewidth]{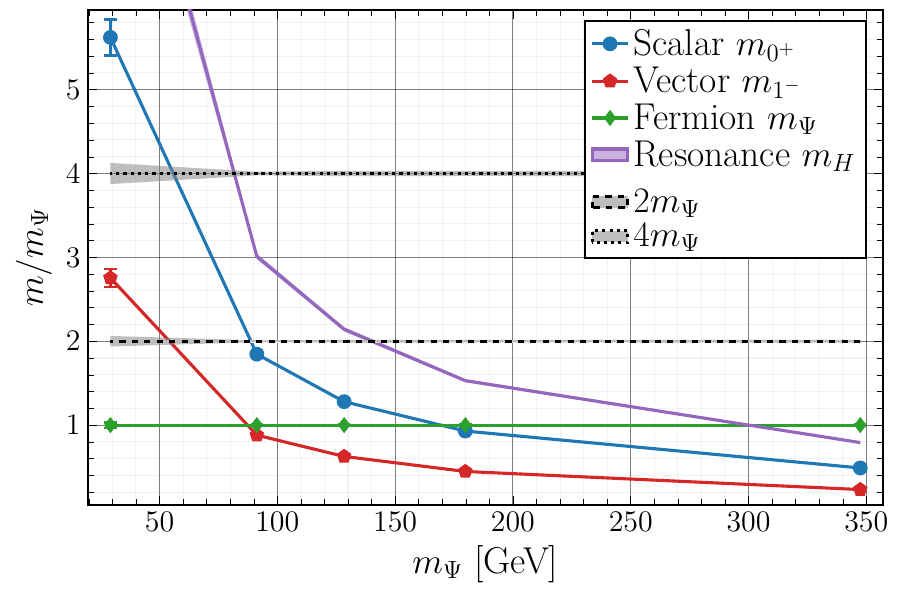}
    \caption{Masses of the FMS observables for the Higgs-like resonance set (set~2). For all fermion masses, the $0^+$~mass is above the~$2m_{1^-}$ threshold which is in agreement with the behavior observed in the gauge-scalar subsystem. Therefore, we also plot the mass of the Higgs-like resonance from~\cite{Jenny:2022atm}.}
    \label{fig:s2}
\end{figure}

\begin{figure}
    \centering
    \includegraphics[width=0.49\linewidth]{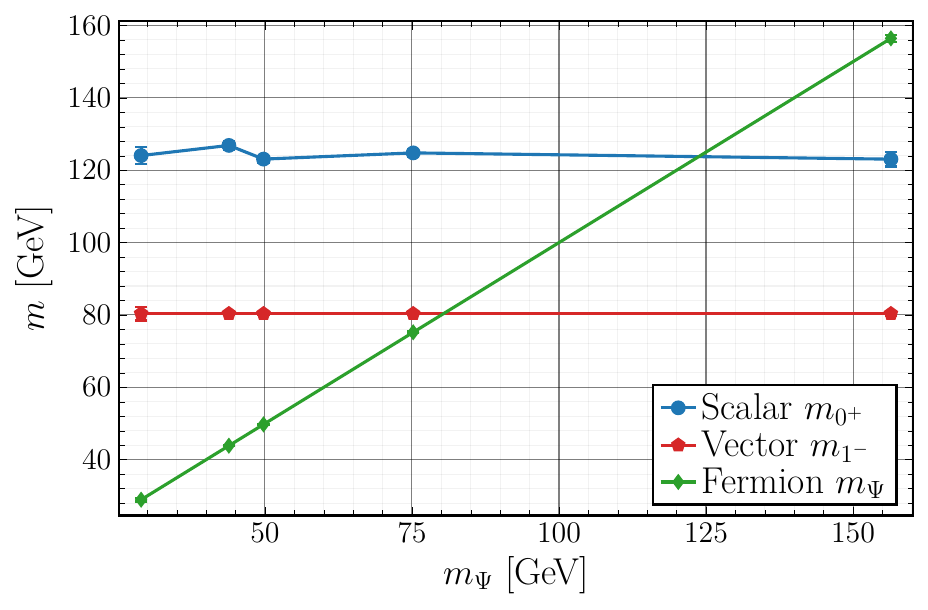} \hfill
    \includegraphics[width=0.49\linewidth]{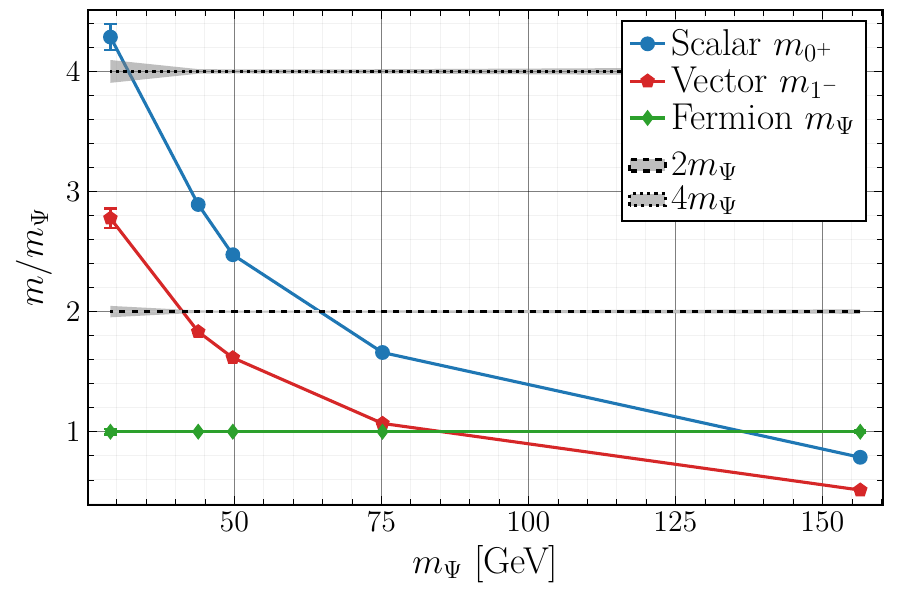}
    \caption{Spectrum for set~3 (weak coupling). Here, a similar behavior compared to set~1 is found with the scalar mass being now clearly below the~$2m_{1^-}$ threshold and closer to its experimentally measured value.}
    \label{fig:s3}
\end{figure}

Figures~\ref{fig:s1} to~\ref{fig:s3} show the energy spectra for all three sets expressed both in physical units and as ratios to the physical fermion mass. For all sets, we find that the gauge-scalar dynamics, i.e., the scalar singlet and vector triplet masses, are almost completely unaffected by the introduction of dynamical fermions. As a result, the system can be controlled fully by the fermion mass and therefore the corresponding lattice parameter. Additionally, since the unphysical phase transition (see appendix~\ref{app:pt}) happens only for smaller fermion masses, all possible decay channels can be accessed by simply adjusting the bare fermion mass parameter~$am_F$.

Although the gauge-scalar dynamics are almost completely unaffected, they differ for the different sets as expected. For set~1, see figure~\ref{fig:s1}, the scalar singlet mass is compatible with/slightly below the two vector triplet threshold. Even though our results do not confirm that the scalar singlet state is stable, they are compatible with simulations of the gauge-scalar system~\cite{Jenny:2022atm}, which produce a stable Higgs. For set~2, we are not able to determine the mass of the scalar singlet as our results give energy levels above the~$2m_{1^-}$ threshold, see figure~\ref{fig:s2}. This is again consistent with gauge-scalar simulations~\cite{Jenny:2022atm}, where the mass in the $0^+$~channel was found to be a resonance. For set~3 (see figure~\ref{fig:s3}), the mass of the scalar is clearly below the threshold and comparable with the experimental value for the Higgs mass.

That the scalar singlet mass for both set~1 and set~2 is close to the~$2m_{1^-}$ threshold is also reflected by the masses of the diagonal correlators in the scalar singlet channel before the EVP analysis. When extracting the masses of the diagonal correlators, we find that the two-W operator, i.e.,~$\mathcal{O}^{{\bf b}, \nu}_{1^-} \, \mathcal{O}^{{\bf b}, \nu}_{1^-}$, gives the smallest energy level. Therefore, neglecting this operator in the EVP results in higher energy levels that do not represent the ground state.

Ultimately, the spectra of the FMS observables demonstrate that all results align with the FMS predictions. The masses of the Higgs and W/Z~bosons are independent of the inclusion of leptons and the leptons themselves can be chosen to be arbitrarily light. This behavior gives rise to a substantial mass defect,  which is incompatible with standard perturbation theory, but in line with the predictions of APT~\cite{Afferrante:2020fhd,Frohlich:1980gj}.

\subsection{Energy spectra of the leptonium channels}
\label{sec:lept}

As a secondary objective, we are able to extract masses for various leptonium channels. We focus on four channels: pseudo-scalar, scalar, vector, and axial-vector; their quantum numbers are given in table~\ref{tab:channels}. For all channels, the lowest energy level determined from the EVP is consistent with the energy level of the operator defined in equation~\eqref{eq:lept}. Therefore, the EVP described in section~\ref{sec:spec} was not necessary to determine the masses of the leptonia.

Compared to other channels, the statistical errors of the leptonium correlators are very small, typically at sub-percent level. However, given the fact that the resulting masses can be quite large compared to the ones of the FMS observables, systematic uncertainties become more relevant. The reasons for this are twofold. First, the correlators span over several orders of magnitude. We have tried to mitigate this by using weighted fits which proved to be somewhat helpful. Second and more severe, the channels are plagued by excited state contaminations making it impossible to identify a clear plateau in the effective mass. While utilizing double cosh fits has been advantageous in this regard, the fitting routines occasionally fail to converge for some or all bootstrap samples. These points are then disregarded in the infinite volume fits, see appendix~\ref{app:ivfits}. For the scalar and axial-vector channels, an additional constant is required increasing the number of fit parameters making the fitting procedure less reliable.

\begin{figure}
    \centering
    \includegraphics[width=0.49\linewidth]{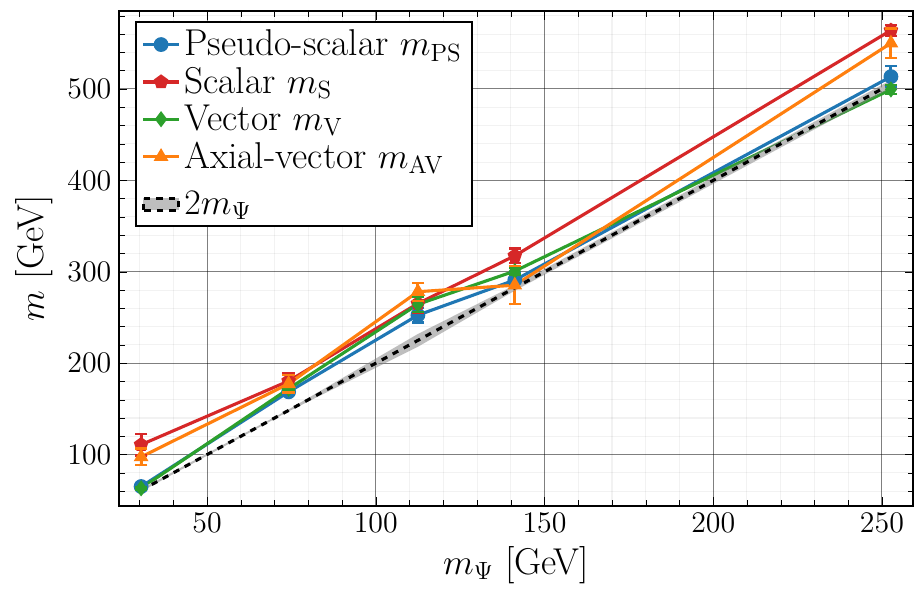} \hfill
    \includegraphics[width=0.49\linewidth]{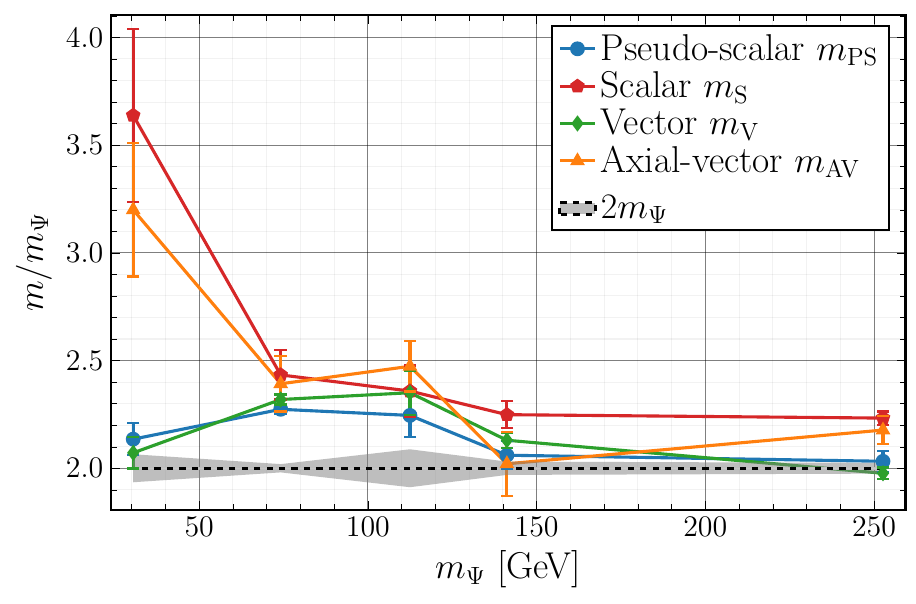} \\
    \includegraphics[width=0.49\linewidth]{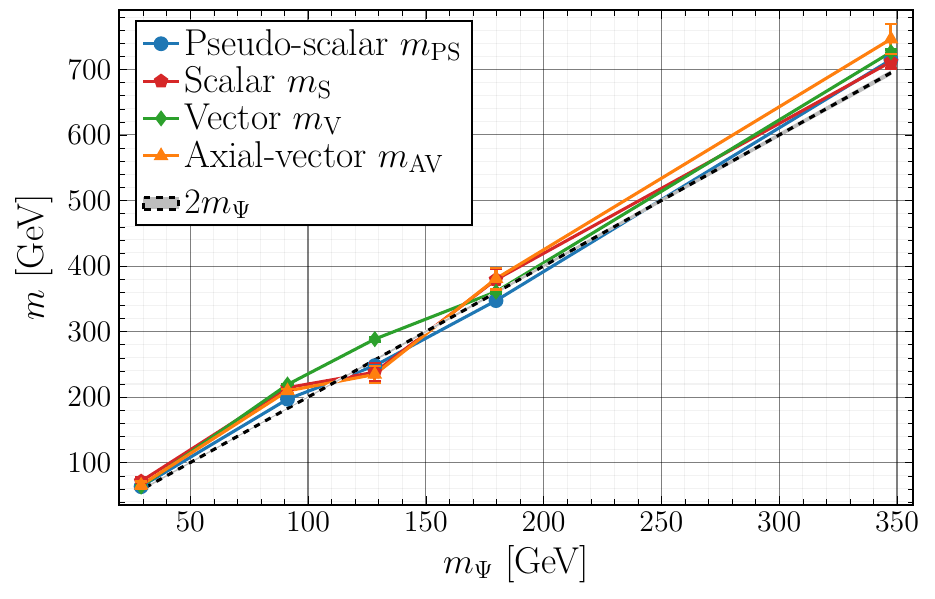} \hfill
    \includegraphics[width=0.49\linewidth]{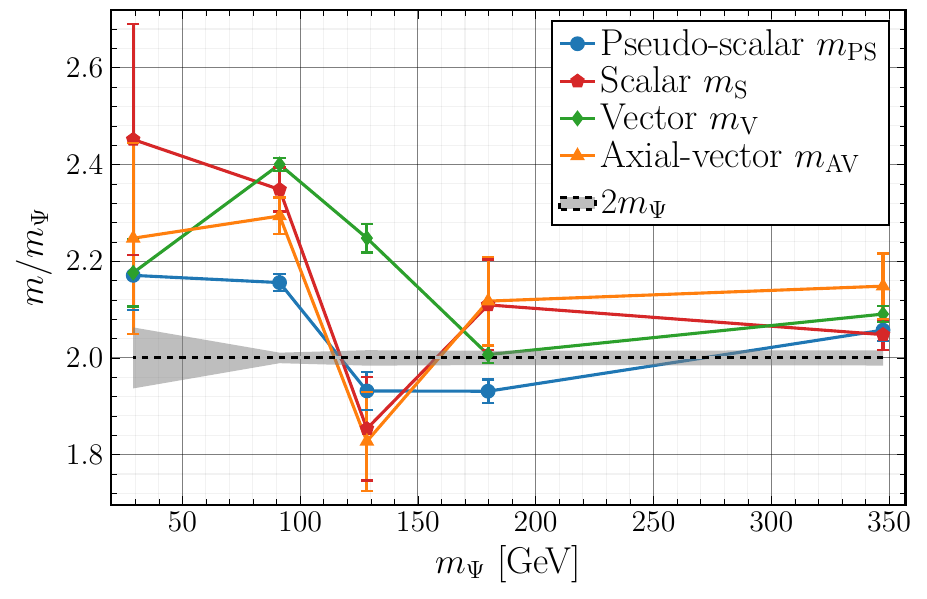} \\
    \includegraphics[width=0.49\linewidth]{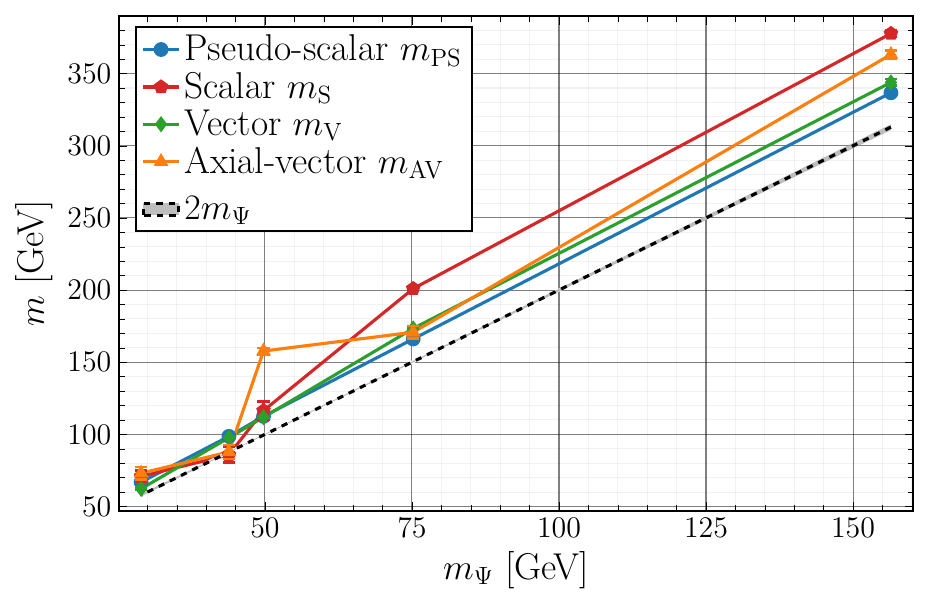} \hfill
    \includegraphics[width=0.49\linewidth]{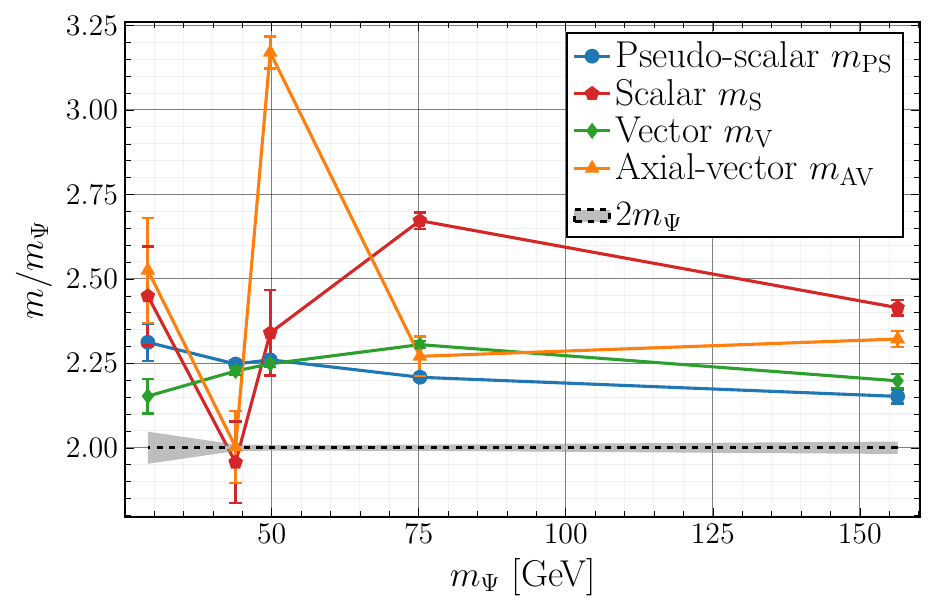} \\
    \caption{Leptonium masses (left) and their ratio with~$m_\Psi$ (right) as a function of the physical fermion mass for all parameter sets (set~1: top, set~2: middle, set~3: bottom). For every set, the four channels listed in table~\ref{tab:channels} are depicted. The plots only include statistical errors; refer to the main text for a discussion on systematic errors. Almost all values are above the~$2m_\Psi$ threshold within errors. The only exception are two points for set~2, however, this deviation is attributed to possible systematic errors and/or statistical fluctuations. The numerical values in the plots on the left can be found in table~\ref{tab:data_iv}.}
    \label{fig:lept}
\end{figure}

Figure~\ref{fig:lept} depicts our results for the masses in all leptonium channels. The errors are statistical only, however, due to the fact that the systematic errors carry over from the energy levels at finite lattice extent to the infinite volume fits, parts of the systematic uncertainties are also included. Overall, we find the leptonium masses to be above the threshold of two physical fermions, i.e., above~$2m_\Psi$. This allows us to draw two major conclusions. First, all considered states in the leptonium channels are unstable. This is inline with the FMS picture, which predicts no additional states besides the Higgs, W/Z~boson, and leptons. Second, all leptonium states are compatible with scattering states of two physical leptons. This hints towards additional Higgs contributions to operators that do not include scalar fields, which is again supported by the FMS mechanism. Note, however, that a dedicated L\"uscher analysis~\cite{Luscher:1986pf} is required to confirm this.

\section{Conclusions}
\label{sec:conclusions}

We studied a proxy to the lepton-weak sector of the Standard Model using an SU(2)~Yang-Mills theory with scalars and dynamical fermions. This system allows to compare non-perturbative predictions from lattice simulations with results from APT. In this first study of the ground state spectrum, we established masses of the physical Higgs, W/Z~bosons, and the leptons with their matching mass-degenerate neutrinos in systems where the fundamental scalar acquires a non-zero vacuum expectation value in minimal 't~Hooft-Landau gauge, thus exhibiting a BEH effect. We assumed that discretization artifacts are small, which is supported by studies in the purely bosonic case~\cite{Maas:2014pba} and by the fact that our three ensembles are consistent despite a factor of almost two in lattice spacing. We show explicitly that we are able to control the statistical and systematic uncertainties associated with finite volume effects and fitting.

In this dynamical setup we identified the bare parameters of the action that correspond to the BEH phase with a finite Higgs vacuum expectation value. In contrast to perturbation theory, APT predicts that the fermionic bound state, consisting of a fundamental scalar and a fundamental lepton, can have arbitrarily light mass. This is the case, even if the physical Higgs, the state consisting of two fundamental scalars, is orders of magnitudes heavier than the physical lepton. We were able to confirm this significant mass defect in our non-perturbative calculations. Furthermore, the reduction of the physical fermion mass left the gauge-scalar dynamics, and correspondingly the W-Higgs mass hierarchy, unaffected. We thereby provide further evidence that the one-to-one correspondence of the W-Higgs mass ratio (with the W always lighter than the Higgs) to having a BEH effect in Landau gauge -- found in all investigations of the gauge-scalar system~\cite{Maas:2017wzi,Maas:2023emb} -- is preserved when fermions are added.

APT further predicts that the composite operators correspond to the actual asymptotic states. Our calculations support this notion, by showing that different bases in an EVP or GEVP analysis do not change the ground state masses in the case where the Higgs is stable. Moreover, no additional stable states have been found, however, the number of tested channels is not exhaustive. We investigated different leptonium channels which are all consistent with scattering states of two physical leptons, again in agreement with predictions.

To further study how to describe the effects of the composite asymptotic state structure, many observables are accessible as next steps. Among these are spectral densities and cross sections. In particular, we can compare predictions (and postdictions) obtained from APT and standard perturbative calculations for such quantities to non-perturbative, first-principles lattice results~\cite{Martins:2026eni}, along the lines of~\cite{Maas:2020kda,Jenny:2022atm}. This will make it possible to assess for which dynamical quantities the use of APT is mandatory and allow us to explore where possible experimental signatures can be expected to be observed.

In this study, only gauged (left-handed) fermions were relevant due to the vanishing Yukawa couplings. Studying the excited state structure in systems with a BEH effect with finite Yukawa couplings will present a way to test the possibility that lepton and quark generations can be explained as excited state structures~\cite{Egger:2017tkd,Greensite:2020lmh}, where masses and mixing are determined over the matrix containing the Yukawa couplings. Future work will cover these higher dimensional systems.

\acknowledgments

We would like to thank Yannick Dengler and Fabian Zierler for helpful discussions. The computational results have been achieved using the Austrian Scientific Computing~(ASC5) infrastructure and the Graz Scientific Cluster~(GSC1).  This research was funded in whole or in part by the Austrian Science Fund~(FWF) [\href{https://www.fwf.ac.at/en/research-radar/10.55776/PAT6443923}{10.55776/PAT6443923}]. The authors acknowledge support from the European Union~(EU) COMETA COST Action~(\href{https://www.cost.eu/actions/CA22130}{CA22130}).

{\bf Data availability statement} -- All data necessary to replicate the EVP analysis and the complete set of calculated ground state energy levels can be downloaded from~\cite{data_release}.

{\bf Code availability statement} -- The version of the \texttt{HiRep} code that was used to produce all configurations and perform all measurements for this paper will be made public in the future.

{\bf Use of generative AI} -- We acknowledge the use of generative AI to support coding. All generated code was independently reviewed, tested, and validated by the authors.

\appendix

\section{Unphysical behavior of the fermion correlator}
\label{app:pt}

\begin{figure}
    \centering
    \includegraphics[width=0.49\linewidth]{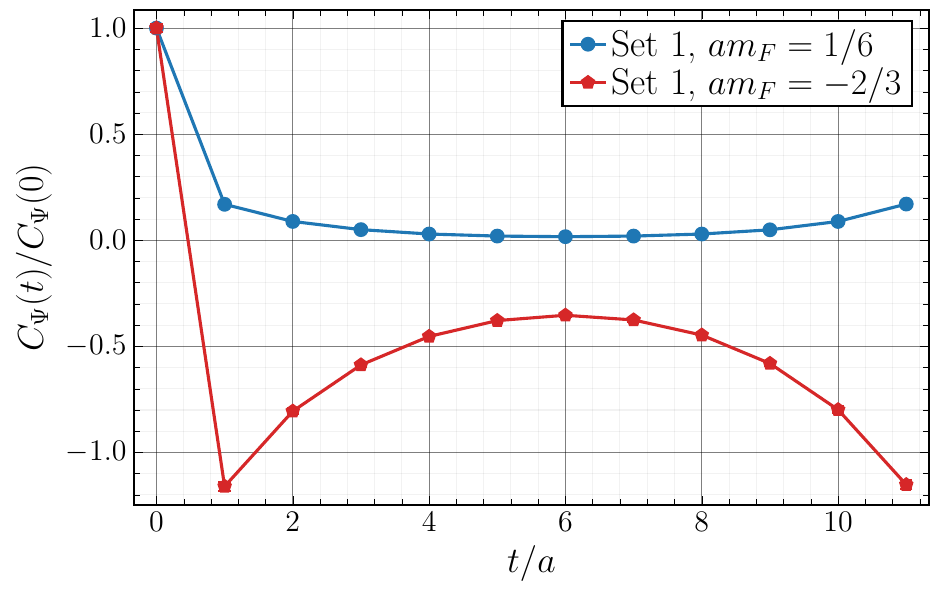} \hfill \includegraphics[width=0.49\linewidth]{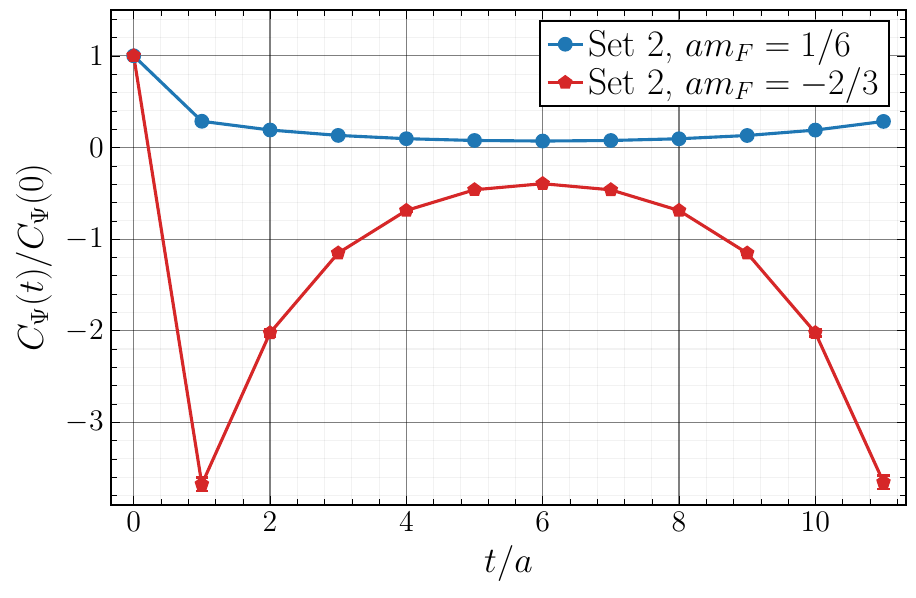}
    \caption{Fermion bound state correlator as a function of Euclidean time for two different values of the bare fermion mass~$am_F$ for set~1 (left) and set~2 (right). The correlators for~$am_F=1/6$ exhibit the expected physical behavior, exponential decay and monotonic decrease. The correlators for~$am_F=-2/3$ show an unphysical behavior, indicating a transition into an unphysical region of the parameter space for smaller values of~$am_F$.}
    \label{fig:corr}
\end{figure}

During a first exploration of the parameter space using smaller lattice sizes, lower values for the bare fermion mass parameter~$am_F$ compared to the ones listed in table~\ref{tab:params} were simulated as well. These simulations resulted in fermion correlators with an unphysical behavior beyond some critical value~$am_F^{\textrm{crit}}$ for all sets discussed in this work. Examples for the behavior of the fermion correlator in the physical and unphysical phase are presented in figure~\ref{fig:corr}. In the physical region, the fermion correlator shows the typical exponential decay and is falling monotonically towards the middle of the lattice. Beyond~$am_F^{\textrm{crit}}$, the correlator becomes negative for all~$t>0$ and rises monotonically towards~$L_t/2$.

The transition into the unphysical region of parameter space does not affect the results presented in the main part of this paper, since all simulation points for all sets are accessible before transitioning into the unphysical phase. The critical values for the bare fermion mass parameter differ for all three sets, which indicates that the gauge-scalar part of theory has an effect on the location of this transition. Due to the change in the fermion correlator, we assume that the unphysical behavior is connected to the loss of reflection positivity. Furthermore, we believe that the location of the transition is influenced by the scalar fields. For a theory without scalar fields,~$am_F^{\textrm{crit}} = -1.0$~\cite{Montvay:1994cy}; thus, adding scalars seems to reduce this value, but only slightly.

\section{Pseudo-scalar decay constant}
\label{app:fps}

Besides the observables discussed in section~\ref{sec:spec}, we are able to extract the pseudo-scalar decay constant~$f_{\textrm{PS}}$ without much additional effort. For calculating~$f_{\textrm{PS}}$, we used two different methods.  First, the decay constant is extracted from the amplitude of the $\gamma_0\gamma_5$~leptonium correlator~\cite{Bennett:2019jzz, Dengler:2024maq}. Second, the analogue of the PCAC mass known from QCD~\cite{Hernandez:2019qed, Martins:2024ilw} is utilized. The latter is expected to be valid only in the domain where chiral perturbation theory applies, which is not the case here. We therefore expect it to differ from the former.

To extract the decay constant from the amplitude of the $\gamma_0\gamma_5$~correlator, we use
\begin{equation}
    \lim_{t \rightarrow \infty} C_{\mathcal{O}_{\gamma_0 \gamma_5}}(t) = \frac{\left(f^{\text{bare}}_{\text{PS}}\right)^2 m_{\text{PS}}}{2} \left[ e^{-m_{\text{PS}} t} + e^{-m_{\text{PS}} (L_t-t)} \right].
\end{equation}
Note that~$m_{\text{PS}}$ is determined from a separate quantity, i.e., the pseudo-scalar leptonium correlator~$C_{\mathcal{O}_{\gamma_5}}(t)$. Finally, a multiplicative renormalization factor is required,~$f_{\text{PS}} = Z_A f^{\text{bare}}_{\text{PS}}$. This factor is an artifact of using Wilson fermions; for details, see~\cite{Bennett:2019jzz}. 

The second method starts with calculating the "renormalized quark mass"~$M^R$, which is given by the PCAC mass,~$M^R = Z_A m_{\text{PCAC}}$. According to~\cite{Kulkarni:2022bvh}, the PCAC mass can be calculated via
\begin{equation}
  m_{\text{PCAC}} = \lim_{t \rightarrow \infty} \frac{1}{2} \frac{\partial_t C_{\mathcal{O}_{\gamma_0 \gamma_5}, \mathcal{O}_{\gamma_5}}(t)}{C_{\mathcal{O}_{\gamma_5}}(t)},
\end{equation}
where~$C_{\mathcal{O}_{\gamma_0 \gamma_5}, \mathcal{O}_{\gamma_5}}(t)$ is a mixed leptonium correlator. The bare pseudo-scalar decay constant is then given by
\begin{equation}
  f_{\text{PS}}^{\text{bare}} = \frac{2 m_{\text{PCAC}} \langle 0 \vert \mathcal{O}_{\gamma_5} \vert \text{PS} \rangle}{m_{\text{PS}}^2},
\end{equation}
where~$\langle 0 \vert \mathcal{O}_{\gamma_5} \vert \text{PS} \rangle$ is related to the amplitude of~$C_{\mathcal{O}_{\gamma_5}}(t)$.

\begin{figure}
    \centering
    \includegraphics[width=0.49\linewidth]{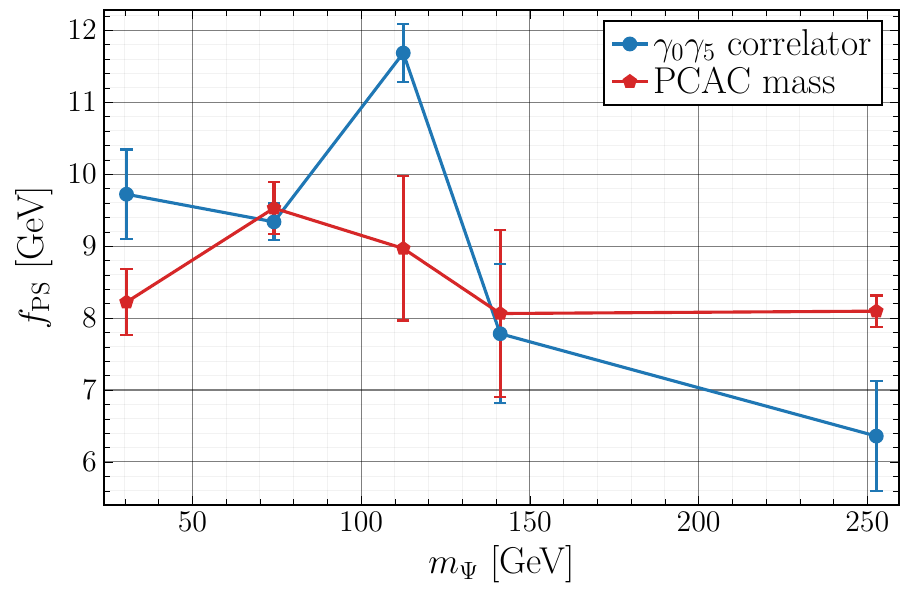} \hfill 
    \includegraphics[width=0.49\linewidth]{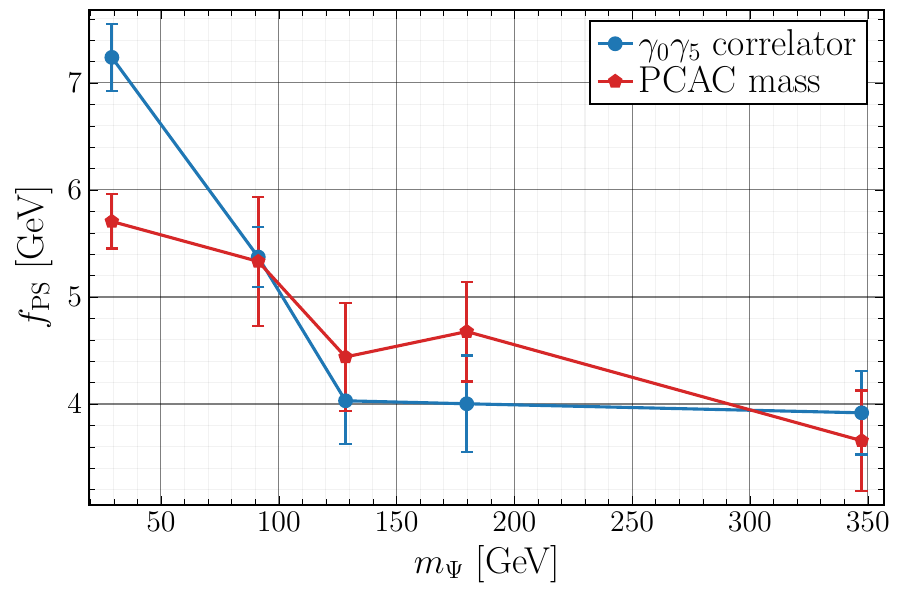} \\
    \includegraphics[width=0.49\linewidth]{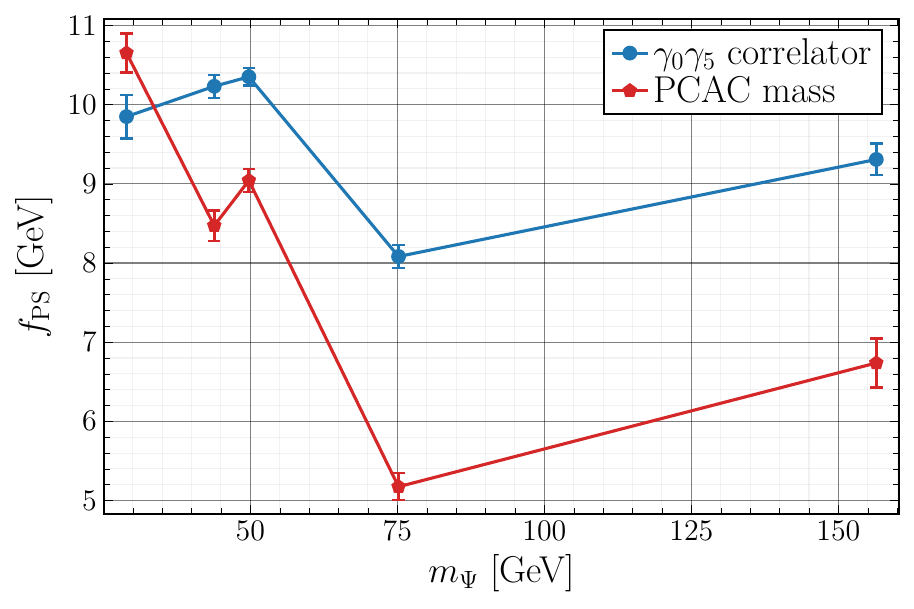}
    \caption{Pseudo-scalar decay constant as a function of the physical fermion mass for the three parameter sets (set~1: top left, set~2: top right, set~3: bottom). The decay constant is determined in two ways: once from the amplitude of the $\gamma_0\gamma_5$~correlator and once from the PCAC mass.}
    \label{fig:fps}
\end{figure}

Figure~\ref{fig:fps} shows the results for the pseudo-scalar decay constant for all three sets extrapolated via~\eqref{eq:iv}. Both methods yield values of the same order of magnitude, but are quantitatively incompatible. Although the systematic errors discussed in section~\ref{sec:lept} become especially relevant for the calculation of~$f_{\textrm{PS}}$, the results support that our simulations are not affected by QCD-like dynamics. Furthermore, no clear trend for different fermion masses is found. The corresponding numerical values are given in table~\ref{tab:fps}. Respective values for finite lattice extent are included in the data release~\cite{data_release}.

\begin{table}
\centering
\begin{tabular}{| c | c | c || c | c | c || c | c | c |}
    \hline
    Ensemble & $f^{(1)}_{\textrm{PS}}$ & $f^{(2)}_{\textrm{PS}}$ & Ensemble & $f^{(1)}_{\textrm{PS}}$ & $f^{(2)}_{\textrm{PS}}$ & Ensemble & $f^{(1)}_{\textrm{PS}}$ & $f^{(2)}_{\textrm{PS}}$ \\
    \hline\hline
    s1-sp1 & $6.4(8)$ & $8.1(3)$ & s2-sp1 & $3.9(4)$ & $3.7(5)$ & s3-sp1 & $9.3(3)$ & $6.7(4)$ \\
    \hline
    s1-sp2 & $7.8(1.0)$ & $8.1(1.2)$ & s2-sp2 & $4.0(5)$ & $4.7(5)$ & s3-sp2 & $8.1(2)$ & $5.2(2)$ \\
    \hline
    s1-sp3 & $11.7(5)$ & $10.0(1.1)$ & s2-sp3 & $4.0(5)$ & $4.4(6)$ & s3-sp3 & $10.35(12)$ & $9.0(2)$ \\
    \hline
    s1-sp4 & $9.3(3)$ & $9.5(4)$ & s2-sp4 & $5.4(3)$ & $5.3(7)$ & s3-sp4 & $10.2(2)$ & $8.5(2)$ \\
    \hline
    s1-sp5 & $9.7(7)$ & $8.2(5)$ & s2-sp5 & $7.2(4)$ & $5.7(3)$ & s3-sp5 & $9.9(3)$ & $10.7(3)$ \\
    \hline
\end{tabular}
\caption{Infinite volume extrapolated values for the pseudo-scalar decay constant for parameter sets~1,~2, and~3. $f^{(1)}_{\textrm{PS}}$~is extracted from the amplitude of the $\gamma_0\gamma_5$~correlator, while~$f^{(2)}_{\textrm{PS}}$ is determined using the PCAC mass. The table lists the central values together with their statistical uncertainties in GeV. Systematic uncertainties are not included.}
\label{tab:fps}
\end{table}

\section{Comparison to GEVP calculations with extended operator bases}
\label{app:gevp}

Complementary to the EVP results shown in the main part of this work, we present results from a GEVP analysis given by the computation of the correlation functions from the diagonalized correlator matrix~$\mathbf{C}_{ij}$,
\begin{equation}
    \mathbf{C}_{ij}(t) v(t) = \lambda(t) \mathbf{C}_{ij}(t_0) v(t)\,,
\end{equation}
where the eigenvalues~$\lambda(t)$ are improved correlation functions associated with the lowest-lying states in the spectrum that have physical overlap with the operators in the chosen basis.

\subsection{Center-of-mass fermionic operators with point sources}

To remove excited state contaminations in a GEVP, we use center-of-mass operators of the form~$\mathcal{O}_{\mathrm{CMS}} = \mathcal{O}(-\bm{p})\mathcal{O}(\bm{p})$. In the~$0^{+}$ and $1^{-}$~channels, the construction is straightforward. For the fermionic operators, the correlation functions need to be built from the lepton propagator (the analogue to the quark propagator in QCD) evaluated with point sources at~$D^{-1}(x|0)$. To construct both the diagonal ($\bm{q}=\bm{p}$) and off-diagonal ($\bm{q}\neq \bm{p}$) matrix elements, we need correlation functions of the form
\begin{equation}
C_{pq}(t) = \langle \widetilde{\mathcal{O}}_{0^{+}}(t,\bm{p})\overline{\widetilde{\mathcal{O}}}_{\mathrm{\Psi}}(t,-\bm{p})\widetilde{\mathcal{O}}_{\Psi}(0,\bm{q})\widetilde{\mathcal{O}}_{0^+}(0,-\bm{q})\rangle\,.
\end{equation}
We can use translational invariance of the expectation value to move the second occurrence of the fermionic operator to the source location~$(0,\bm{0})$:
\begin{align}
C_{pq}(t) &= \sum_{\bm{x_1},\bm{x_2},\bm{x_3},\bm{x_4}}e^{-\tfrac{2\pi\mathrm{i}}{L}(p(\bm{x_1}-\bm{x_2}) + q(\bm{x_3}-\bm{x_4}))}\langle\mathcal{O}_{0^+}(t,\bm{x_1}) \overline{\mathcal{O}}_{\Psi}(t,\bm{x_2})\mathcal{O}_\Psi(0,\bm{x_3})\mathcal{O}_{0^+}(0,\bm{x_4})\rangle\\
&=\sum_{\bm{x_1'},\bm{x_2'},\bm{x_3},\bm{x_4'}}e^{-\tfrac{2\pi\mathrm{i}}{L}(p(\bm{x_1'}-\bm{x_2'}) - q\bm{x_4'})}\langle\mathcal{O}_{0^+}(t,\bm{x_1'}) \overline{\mathcal{O}}_{\Psi}(t,\bm{x_2'})\mathcal{O}_\Psi(0,\bm{0})\mathcal{O}_{0^+}(0,\bm{x_4'})\rangle.
\end{align}
This correlation function can be approximated on the lattice using the estimator
\begin{align}
\hat{C}_{pq(t)} &= L^3\sum_{\bm{x_1'},\bm{x_2'},\bm{x_4'}}e^{-\tfrac{2\pi\mathrm{i}}{L}(p(\bm{x_1'}-\bm{x_2'}) - q\bm{x_4'})}\mathcal{O}_{0^+}(t,\bm{x_1'}) X^{\dagger}(\bm{x_2'})D^{-1}_{\overline{\psi}\psi}(\bm{x_2'}|0)X(0)\mathcal{O}_{0^+}(0,\bm{x_4'}) \\ &= L^3 \widetilde{\mathcal{O}}_{0^+}(t,\bm{p})\left[\sum_{\bm{x}}e^{-\tfrac{2\pi\mathrm{i}}{L}\bm{x}(-\bm{p})}X^{\dagger}(x)D^{-1}_{\overline{\psi}\psi}(x|0)X(0)\right]\widetilde{\mathcal{O}}_{0^{+}}(t,-\bm{q})\,.
\end{align}
This expression is the product of the projected operators and correlators. Consequently, we can construct the fermionic higher-energy center-of-mass operators from the projected point-source physical fermion propagators in a completely naive way.

\subsection{Interpretation of the results for different operator bases}

Uncertainties quoted in this paper are mostly computed from bootstrap resampling of decorrelated data. A complementary method to estimate uncertainties from autocorrelations is the Gamma method~\cite{Wolff:2003sm}. We further need to be reasonably sure that the chosen operator bases have sufficient overlap with the actual ground states. To check the correctness of the autocorrelations and the choice of operator bases, selected sets where shadow analyzed using a GEVP analysis and the Gamma method, employing the python package \texttt{pyerrors}~\cite{Joswig:2022qfe, Wolff:2003sm, Ramos:2018vgu, Schaefer:2010hu}. 

For the GEVP calculations, we introduce two new operators in the scalar singlet and vector triplet channels~\cite{Jenny:2022atm,Evertz:1985fc}:
\begin{align}
    \mathcal{O}_{0^+,\mathrm{N}} &= \sum_{\mu=1}^3 \mathrm{tr} \left[ \frac{X^\dagger(x)}{\sqrt{\mathrm{det} \, X(x)}} U^\mu(x) \frac{X(x+\hat{\mu})}{\sqrt{\mathrm{det} \, X(x+\hat{\mu})}} \right],\\
    \mathcal{O}^{{\bf a}, \mu}_{1^-,\mathrm{N}} &= -\frac{i}{2} \mathrm{tr} \left[ \tau^{\mathbf{a}} \frac{X^\dagger(x)}{\sqrt{\mathrm{det} \, X(x)}} U^\mu(x) \frac{X(x+\hat{\mu})}{\sqrt{\mathrm{det} \, X(x+\hat{\mu})}} \right].
\end{align}
These proved to be less noisy compared to the operators defined in equations~\eqref{eq:ops1} and~\eqref{eq:ops2}. The operator bases used here are constrained by positive definiteness of the resulting GEVP matrix and depicted in table~\ref{tab:gevpbasis} (to compare with the EVP bases, see table~\ref{tab:evp}). In both cases, four iterations of APE smearing with~$\alpha_{\text{APE}}=0.45$ were used. In contrast to the EVP bases, more excited state structure was captured due to higher momenta center-of-mass operators in all channels. In the vector and fermion channel,~$\bm{p=0}$ operators were omitted due to noisy overlap with the vacuum.

\begin{figure}
    \centering
    \includegraphics[width=0.49\linewidth]{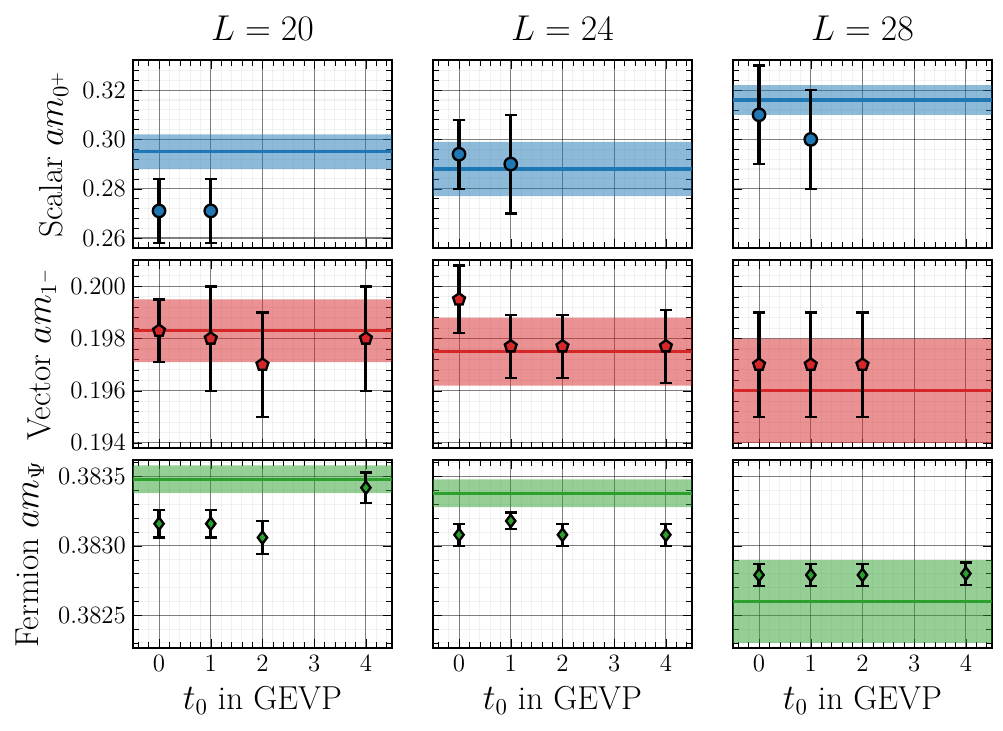} \hfill \includegraphics[width=0.49\linewidth]{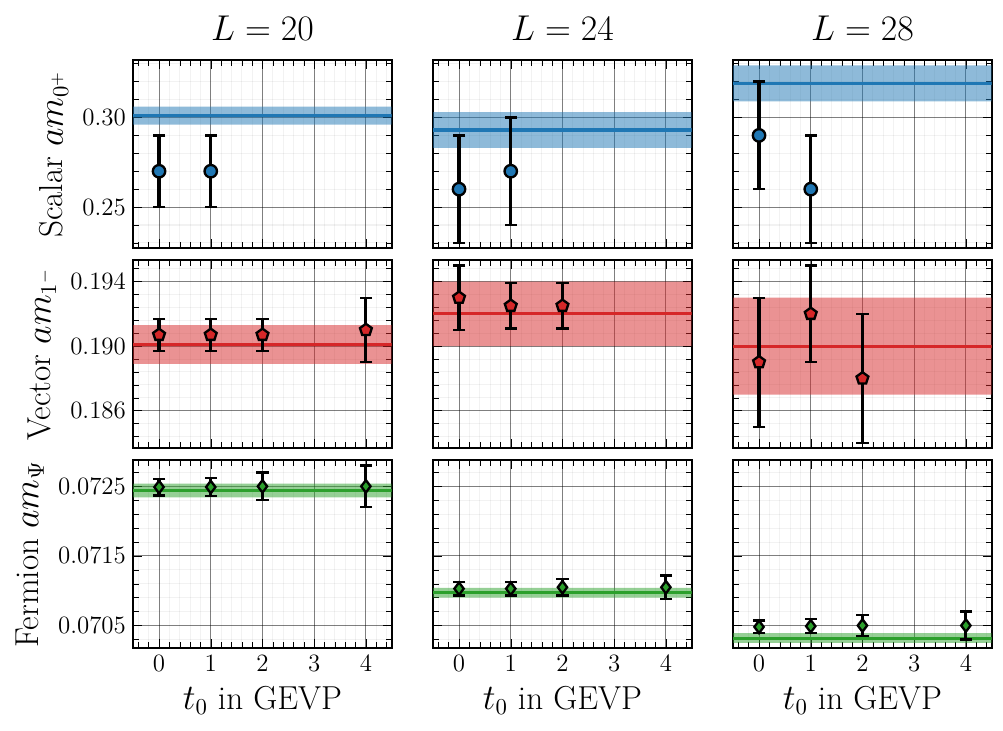}
    \caption{Comparison between EVP and GEVP results for ensembles~s3-sp1 (left) and~s3-sp5 (right). The lines with ribbons correspond to the EVP ground state energy levels, whereas the markers depict the GEVP results for different~$t_0$.}
    \label{fig:gevp}
\end{figure}

The results in figure~\ref{fig:gevp} and in tables~\ref{tab:gevp-s1} and~\ref{tab:gevp-s3} show that while the GEVP gives in some cases slightly lower estimates compared to the EVP, most of the energy levels agree within the quoted errors. This is evidence for the FMS mechanism, implying that the usual LSZ mechanism is at play despite the BEH effect~\cite{Frohlich:1981yi,Maas:2024hcn}, and any gauge-invariant operator with the correct quantum numbers will yield the correct asymptotic ground state. For sets~1 and~2, however, it is important to include the structural two-W operator, because this operator produces the cleanest ground state signal. This has already been seen in the gauge-scalar case~\cite{Jenny:2022atm} and is corroborated for dynamical simulations by both the EVP and GEVP analyses.

\begin{table}
  \centering
  \begin{tabular}{| c || c | c | c |}
  \hline
    Channel & $\,$ Scalar singlet $\,$ & $\,$ Vector triplet $\,$ & $\,$ Fermion doublet $\,$ \\
    \hline\hline
    \multirow{4}{6.9em}{\centering Basis elements} 
        & $\mathcal{O}_{0^+,\mathrm{N}}$ 
        & $\mathcal{O}^{{\bf a}, \mu}_{1^-}$ 
        & $\mathcal{O}^{\mathbf{i}, \alpha}_{\Psi}$  \\
        
        & $\mathrm{tr} \left[P_{\mu\nu}(x)\right]$ 
        & $\mathcal{O}^{{\bf a}, \mu}_{1^-,\mathrm{N}}$ 
        & $\, \mathcal{O}_{0^+}(\bm{\bm{p}}) \, \mathcal{O}^{\mathbf{i}, \alpha}_{\Psi}(\bm{-p})$, $|\bm{p}| = 1,2 \, $  \\
        
        & $\mathcal{O}^{{\bf b}, \nu}_{1^-}(\bm{p=0}) \, \mathcal{O}^{{\bf b}, \nu}_{1^-}(\bm{p=0})$ 
        & $\mathcal{O}_{0^+}(\bm{\bm{p}}) \, \mathcal{O}^{{\bf a}, \mu}_{1^-}(\bm{-p})$, $|\bm{p}| = 1,2$ 
        & \\

        & $\, \mathcal{O}^{{\bf b}, \nu}_{1^-,\mathrm{N}}(\bm{p=0}) \, \mathcal{O}^{{\bf b}, \nu}_{1^-,\mathrm{N}}(\bm{p=0}) \, $ 
        & $\, \mathcal{O}_{0^+,\mathrm{N}}(\bm{\bm{p}}) \, \mathcal{O}^{{\bf a}, \mu}_{1^-,\mathrm{N}}(\bm{-p})$, $|\bm{p}| = 1,2 \, $ 
        & \\[2pt]
    \hline
  \end{tabular}
  \caption{Operator bases used in the GEVP analysis.}
  \label{tab:gevpbasis}
\end{table}

\begin{table}
\centering
\begin{tabular}{| c | c | c || c | c | c || c | c | c |}
\hline
\multirow{2}{1em}{\centering $L$} & \multirow{2}{5em}{\centering Technique} & \multirow{2}{1em}{\centering $t_0$} & \multicolumn{3}{|c||}{s1-sp1} & \multicolumn{3}{|c|}{s1-sp5} \\
\cline{4-9}
   &           &       & $am_{0^+}$ & $am_{1^-}$ & $am_{\Psi}$ & $am_{0^+}$ & $am_{1^-}$ & $am_{\Psi}$ \\
\hline\hline
16 & EVP  &   & 0.51(3)    & 0.261(3)   & 0.785(2)  & 0.44(3)     & 0.212(3)  & 0.1042(5) \\
\hline
16 & GEVP & 0 & 0.42(4)    & 0.263(3)   & 0.7837(6) & 0.35(2)   & 0.213(3)  & 0.1041(9) \\
16 & GEVP & 1 &            & 0.263(3)   & 0.7837(6) &             & 0.214(3)  & 0.1041(9) \\
16 & GEVP & 2 &            & 0.263(3)   & 0.7837(6) &             & 0.212(6)  & 0.1041(10) \\
16 & GEVP & 4 &            &            & 0.7837(6) &             &           & 0.104(2)  \\
\hline
20 & EVP  &   & 0.38(9)    & 0.249(3)   & 0.782(3)  & 0.45(3)     & 0.223(3)  & 0.0937(2) \\
\hline
20 & GEVP & 0 & 0.41(6)    & 0.247(4)   & 0.7821(4) & 0.38(3)     & 0.219(5)  & 0.0935(2) \\
20 & GEVP & 1 & 0.41(5)    & 0.247(4)   & 0.7821(4) & 0.46(2)     & 0.217(5)  & 0.0935(2) \\
20 & GEVP & 2 &            & 0.247(4)   & 0.7821(4) &             & 0.223(3)  & 0.0937(3) \\
20 & GEVP & 4 &            &            & 0.7821(4) &             &           & 0.0933(7) \\
\hline
24 & EVP  &   & 0.43(7)    & 0.251(3)   & 0.7805(7)  & 0.40(5)     & 0.224(3)  & 0.09015(14) \\
\hline
24 & GEVP & 0 & 0.46(6)    & 0.250(4)   & 0.7810(3) & 0.35(4)     & 0.223(4)  & 0.0901(4) \\
24 & GEVP & 1 & 0.48(4)    & 0.249(4)   & 0.7810(3) & 0.36(3)     & 0.221(4)  & 0.0901(4) \\
24 & GEVP & 2 &            & 0.245(8)   & 0.7810(3) &             & 0.221(5)  & 0.0901(4) \\
24 & GEVP & 4 &            &            & 0.7810(3) &             &           & 0.0902(4) \\
\hline
28 & EVP  &   & 0.55(4)    & 0.256(4)  & 0.779(5)  & 0.53(4)     & 0.227(4)  & 0.0886(5)  \\
\hline
28 & GEVP & 0 & 0.38(6)    & 0.257(5)   & 0.7810(3) & 0.44(5)     & 0.224(6)  & 0.08813(10) \\
28 & GEVP & 1 & 0.47(3)    & 0.257(5)   & 0.7810(3) & 0.45(4)     & 0.224(6)  & 0.08813(10) \\
28 & GEVP & 2 &            & 0.257(5)   & 0.7810(4) &             &           & 0.08813(10) \\
28 & GEVP & 4 &            &            & 0.7810(4) &             &           & 0.0882(2) \\
\hline
\end{tabular}
\caption{Comparison between EVP and GEVP results for set~1, ensembles~sp1 and~sp5.}
\label{tab:gevp-s1}
\end{table}

\begin{table}
\centering
\begin{tabular}{| c | c | c || c | c | c || c | c | c |}
\hline
\multirow{2}{1em}{\centering $L$} & \multirow{2}{5em}{\centering Technique} & \multirow{2}{1em}{\centering $t_0$} & \multicolumn{3}{|c||}{s3-sp1} & \multicolumn{3}{|c|}{s3-sp5} \\
\cline{4-9}
 &  & & $am_{0^+}$ & $am_{1^-}$ & $am_{\Psi}$ & $am_{0^+}$ & $am_{1^-}$ & $am_{\Psi}$ \\
\hline\hline
16 & EVP  &   & 0.284(12)  & 0.2023(12)   & 0.38462(13) & & & \\
\hline
16 & GEVP & 0 & 0.267(11)    & 0.2019(12)  & 0.3840(3) & & & \\
16 & GEVP & 1 &                          & 0.201(2)    & 0.3840(3) & & & \\
16 & GEVP & 2 &                          & 0.201(2)    & 0.3840(3) & & & \\
16 & GEVP & 4 &                          &             & 0.3840(3)  & & & \\
\hline
20 & EVP  &   & 0.295(7)                 & 0.1983(12)  & 0.38348(10) & 0.301(5)    & 0.1901(12)  & 0.07244(10) \\
\hline
20 & GEVP & 0 & 0.271(13)    & 0.1983(12)  & 0.38316(10) & 0.27(2) & 0.1907(10)  & 0.07249(12) \\
20 & GEVP & 1 & 0.271(13)    & 0.198(2)    & 0.38316(10) & 0.27(2) & 0.1907(10)  & 0.07249(13) \\
20 & GEVP & 2 &                          & 0.197(2)    & 0.38306(12) &                     & 0.1907(10)  & 0.0725(2) \\
20 & GEVP & 4 &                          & 0.198(2)  & 0.38342(11) &                     & 0.191(2)                & 0.0725(3)  \\
\hline
24 & EVP  &   & 0.288(11)                & 0.1975(13)  & 0.38338(10) & 0.293(10)           & 0.192(2)                & 0.07097(7) \\
\hline
24 & GEVP & 0 & 0.294(14)    & 0.1995(13)  & 0.38308(8) & 0.26(3) & 0.193(2)                & 0.07103(10) \\
24 & GEVP & 1 & 0.29(2)      & 0.1977(12)  & 0.38318(6) & 0.27(3) & 0.1925(14)              & 0.07103(10) \\
24 & GEVP & 2 &                          & 0.1977(12)  & 0.38308(8) &                     & 0.1925(14)              & 0.07105(12) \\
24 & GEVP & 4 &                          & 0.1977(14)  & 0.38308(8) &                     &                         & 0.07105(17) \\
\hline
28 & EVP  &   & 0.316(6)                 & 0.196(2)    & 0.3826(3) & 0.319(10)   & 0.190(3)                & 0.07032(7) \\
\hline
28 & GEVP & 0 & 0.31(2)      & 0.197(2)    & 0.38279(8) & 0.29(3) & 0.189(4)                & 0.07048(9) \\
28 & GEVP & 1 & 0.30(2)      & 0.197(2)    & 0.38279(8) & 0.26(3) & 0.192(3)                & 0.07049(10) \\
28 & GEVP & 2 &                          & 0.197(2)    & 0.38279(8) &                     & 0.188(4)    & 0.07050(15) \\
28 & GEVP & 4 &                          &       & 0.38280(8) &                     &                         & 0.0705(2) \\
\hline
32 & EVP  &   & 0.317(9)                 & 0.200(3)    & 0.3832(4)  & 0.317(12)           & 0.195(6)                & 0.07002(8)  \\
\hline
32 & GEVP & 0 & 0.32(2)      & 0.200(3)    & 0.38259(10) & 0.30(4) & 0.193(6)                & 0.07007(10) \\
32 & GEVP & 1 & 0.32(2)      & 0.200(3)    & 0.38259(10) & 0.29(3) & 0.192(6)    & 0.0701(2) \\
32 & GEVP & 2 & 0.30(2)      & 0.200(3)    & 0.38259(10) &                     & 0.190(7)                & 0.0700(3) \\
32 & GEVP & 4 &                          & 0.200(3)    & 0.38259(10) &                     &                         & 0.0700(5) \\
\hline
\end{tabular}
\caption{Comparison between EVP and GEVP results for set~3, ensembles~sp1 and~sp5.}
\label{tab:gevp-s3}
\end{table}

\section{Infinite volume behavior}
\label{app:ivfits}

As described in the main text, the infinite volume masses are extracted from fits of the form~\eqref{eq:iv}. Here, we would like to show some example plots and argue based on them that the volume behavior is very mild and that it is most often dominated by a constant contribution for larger lattice extent. To achieve satisfactory fits, we have been very careful in choosing the points to be included in the fitting procedure. If energy levels for finite lattice extent are not approximately Gaussian distributed or if their respective fits failed, they are automatically excluded. Additionally, since we use Bayesian inference, the included points are weighted according to their errors.

\begin{figure}
    \centering
    \includegraphics[width=0.49\linewidth]{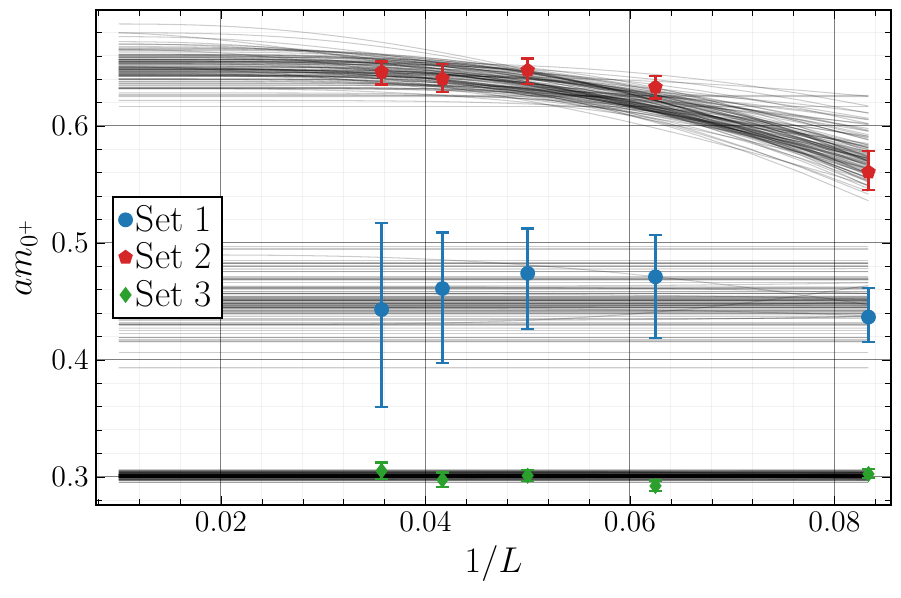} \hfill 
    \includegraphics[width=0.49\linewidth]{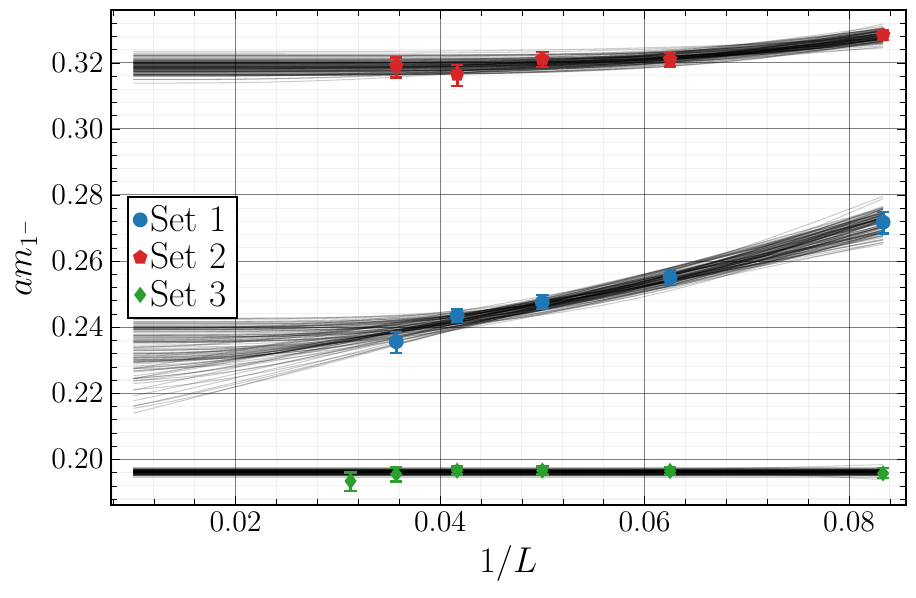}
    \caption{Examples for the volume behavior of the scalar and the vector channels for all sets at~sp3. For almost all channels and sets a weak volume dependence is found. The fit function in equation~\eqref{eq:iv} effectively captures occasional instances of a stronger volume dependence. The lines depict~$100$ random fit results out of~$4000$ in total. Only data points used during the fitting procedure are shown.}
    \label{fig:ivfits-gs}
\end{figure}

For the purely bosonic observables, some example plots are presented in figure~\ref{fig:ivfits-gs}. The volume behavior for most masses are clearly dominated by a constant at large~$L$. One exception in the vector channel is found for set~1. In this particular case, the infinite volume fit works fine and the resulting infinite volume mass is compatible with the energy level of the largest lattice size.

\begin{figure}
    \centering
    \includegraphics[width=0.49\linewidth]{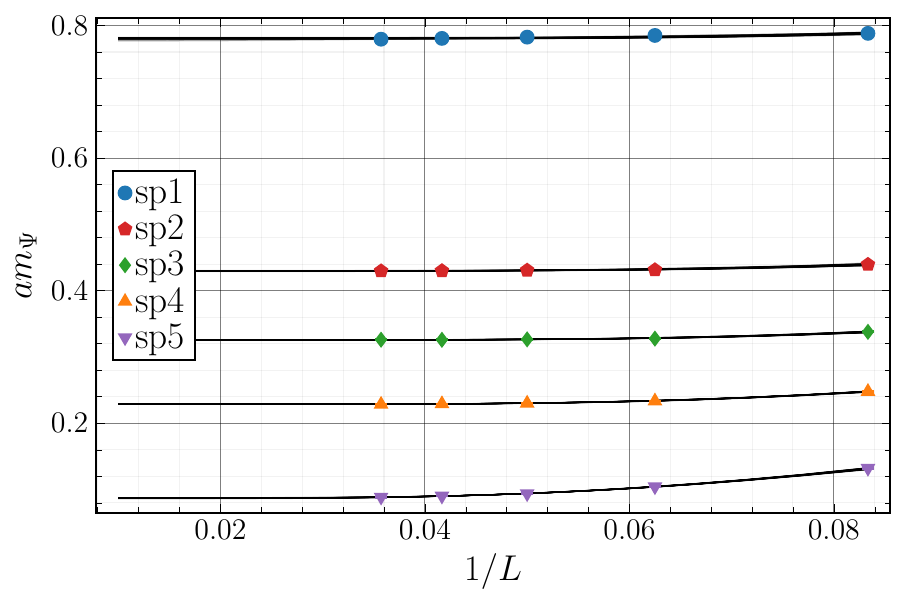} \hfill 
    \includegraphics[width=0.49\linewidth]{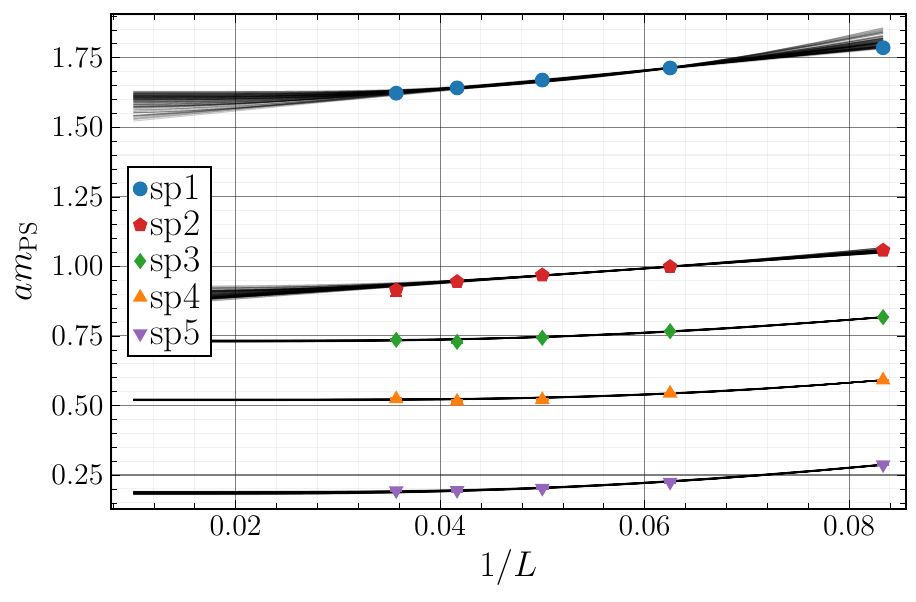} \\
    \includegraphics[width=0.49\linewidth]{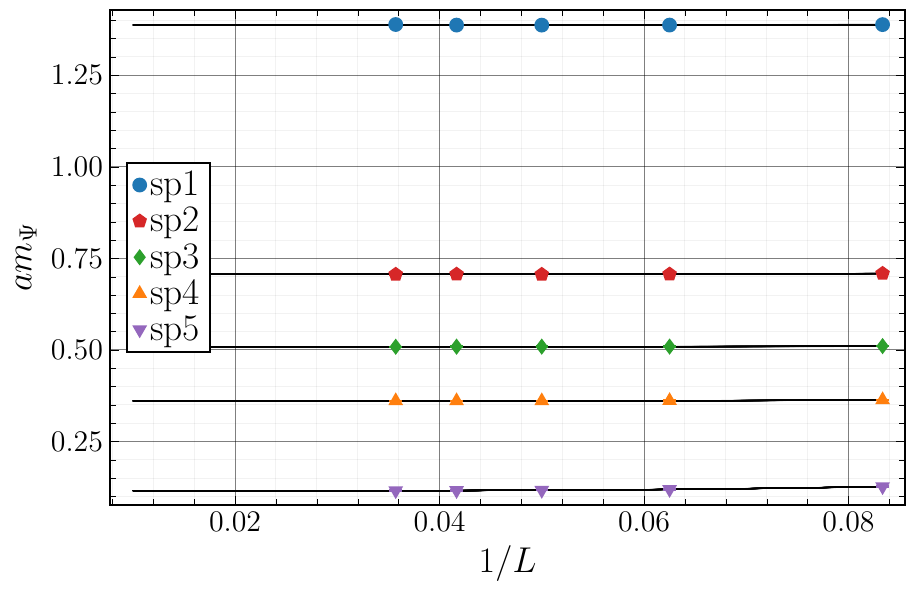} \hfill 
    \includegraphics[width=0.49\linewidth]{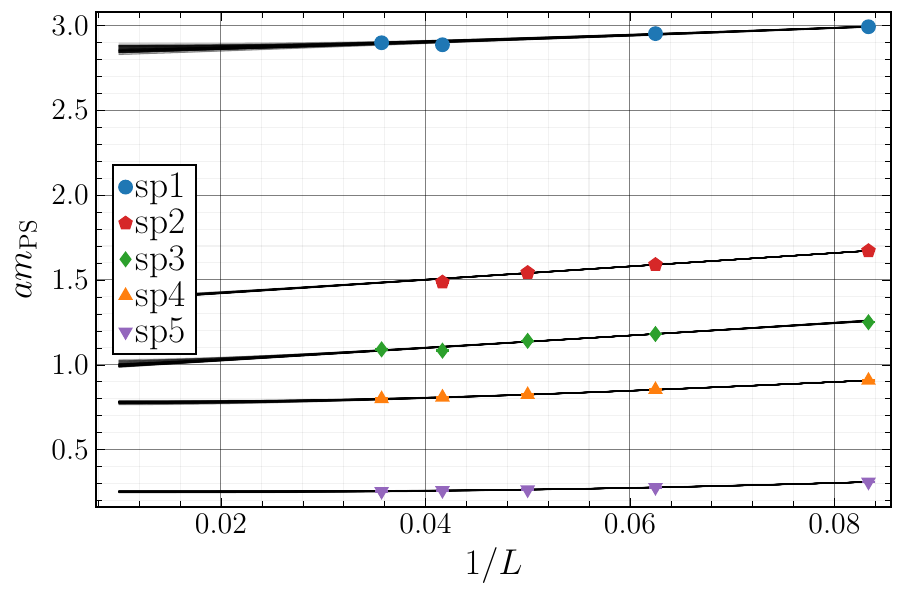} \\
    \includegraphics[width=0.49\linewidth]{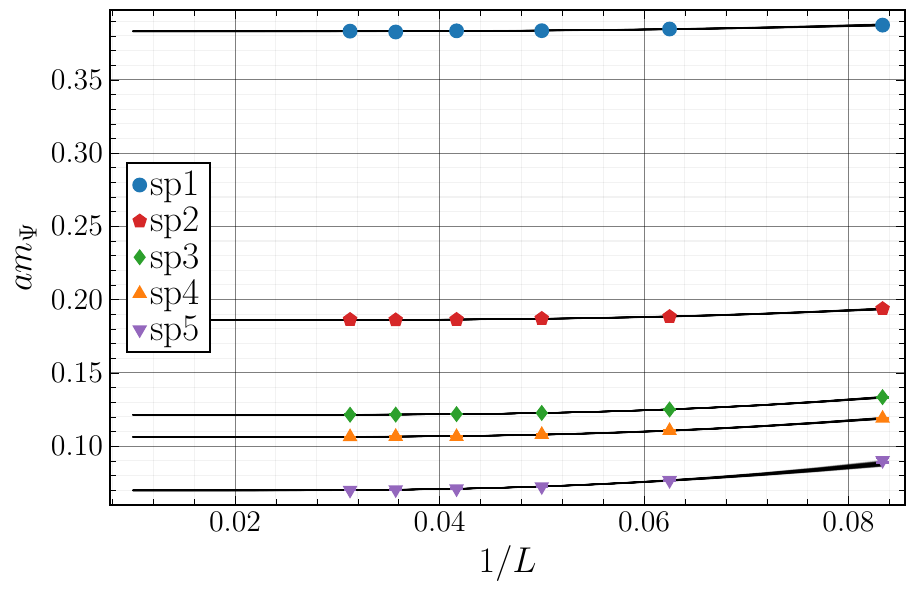} \hfill 
    \includegraphics[width=0.49\linewidth]{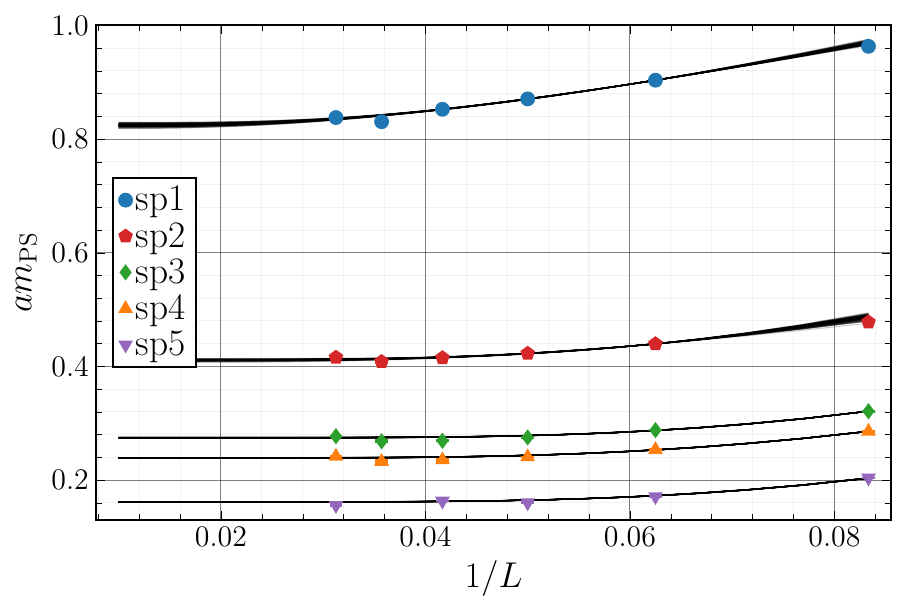}
    \caption{Volume behavior of the fermion bound state and the pseudo-scalar leptonium for all sets (top: set~1, middle: set~2, bottom: set~3). For both channels a weak volume dependence is found. The lines depict~$100$ random fit results out of~$4000$ in total. Only data points used during the fitting procedure are shown; those where the correlator fits failed to converge are omitted.}
    \label{fig:ivfits-f}
\end{figure}

The left plots in figure~\ref{fig:ivfits-f} show some example fits for the fermion mass. All fits are dominated by constants showing an overall weak dependence on the lattice size. For the leptonium channels and the pseudo-scalar decay constant, the situation is similar to the one depicted on the right-hand side of figure~\ref{fig:ivfits-f}. Here, the volume dependence at larger~$L$ is somewhat more noticeable and the systematic errors related to the fitting of the correlators become apparent. Nevertheless, the fitting function in~\eqref{eq:iv} still proofed to be a good choice to extract the infinite volume leptonium masses and the pseudo-scalar decay constant.

\section{Lattice data}
\label{app:data}

In the following tables, lattice data for all ensembles can be found. The tables include the lattice volume~$L^4$, the number of configurations that are used for the analysis~$N_{\textrm{cnfg}}$, the number of bins~$N_{\textrm{bin}}$, the average plaquette~$\langle P \rangle$, the average scalar length~$\langle \sqrt{\phi^\dagger \phi} \rangle$, the vacuum expectation value~$av$, and masses for the FMS observables in lattice units. The data presented in these tables, additional data for all leptonium ground state masses, and values for the pseudo-scalar decay constant can be found in the dedicated data release~\cite{data_release}.

The number of configurations included in the analysis~$N_{\textrm{cnfg}}$ is determined by taking all generated configurations, skipping~$L/2$, and removing enough configurations until the system is thermalized. Thus, the total number of generated configurations is given by~$(N_{\textrm{cnfg}} + N_{\textrm{therm}})\cdot L/2$, where~$N_{\textrm{therm}} \sim \mathcal{O}(100)$. For measuring the vacuum expectation value, gauge fixing is required. To keep the numerical effort at a minimum, only~$1000$ thermalized configurations are used to determine~$av$. For all other quantities shown in the tables,~$N_{\textrm{cnfg}}$ configurations are used in the analysis.

\begin{table}
\centering
\begin{tabular}{| c | c | c | c || c | c | c || c | c | c |}
    \hline
    Ensemble & $L$ & $N_{\textrm{cnfg}}$ & $N_{\textrm{bin}}$ & $\langle P \rangle$ & $\langle \sqrt{\phi^\dagger \phi} \rangle$ & $a v$ &$am_{0^+}$ & $am_{1^-}$ & $am_{\Psi}$ \\
    \hline\hline
    s1-sp1 & $12$ & $19860$ & $4$ & $0.715437(8)$ & $0.599459(10)$ & $0.2013(4)$ & $0.50(3)$ & $0.285(3)$ & $0.7881(11)$ \\
    \hline
    s1-sp1 & $16$ & $18885$ & $2$ & $0.715426(5)$ & $0.599532(6)$ & $0.2003(2)$ & $0.51(3)$ & $0.261(3)$ & $0.785(2)$ \\
    \hline
    s1-sp1 & $20$ & $19900$ & $2$ & $0.715425(3)$ & $0.599543(4)$ & $0.19942(12)$ & $0.38(9)$ & $0.249(3)$ & $0.782(3)$ \\
    \hline
    s1-sp1 & $24$ & $20655$ & $2$ & $0.715429(2)$ & $0.599548(3)$ & $0.19903(9)$ & $0.43(7)$ & $0.251(3)$ & $0.7805(7)$ \\
    \hline
    s1-sp1 & $28$ & $11255$ & $2$ & $0.715431(2)$ & $0.599547(3)$ & $0.19872(6)$ & $0.55(4)$ & $0.256(4)$ & $0.779(5)$ \\
    \hline\hline
    s1-sp2 & $12$ & $19860$ & $4$ & $0.717701(8)$ & $0.599853(11)$ & $0.2050(4)$ & $0.46(3)$ & $0.277(4)$ & $0.4394(10)$ \\
    \hline
    s1-sp2 & $16$ & $19885$ & $4$ & $0.717700(4)$ & $0.599914(6)$ & $0.2038(3)$ & $0.46(3)$ & $0.255(3)$ & $0.4312(10)$ \\
    \hline
    s1-sp2 & $20$ & $19900$ & $2$ & $0.717717(3)$ & $0.599927(4)$ & $0.20334(11)$ & $0.36(8)$ & $0.249(3)$ & $0.4307(3)$ \\
    \hline
    s1-sp2 & $24$ & $19910$ & $2$ & $0.717716(2)$ & $0.599929(3)$ & $0.20291(8)$ & $0.41(6)$ & $0.249(3)$ & $0.4299(4)$ \\
    \hline
    s1-sp2 & $28$ & $10132$ & $2$ & $0.717715(2)$ & $0.599930(3)$ & $0.20246(6)$ & $0.48(8)$ & $0.245(4)$ & $0.4294(10)$ \\
    \hline\hline
    s1-sp3 & $12$ & $19860$ & $4$ & $0.718644(8)$ & $0.600007(10)$ & $0.2071(4)$ & $0.44(3)$ & $0.272(4)$ & $0.3378(6)$ \\
    \hline
    s1-sp3 & $16$ & $19885$ & $4$ & $0.718667(4)$ & $0.600075(6)$ & $0.2056(2)$ & $0.46(7)$ & $0.255(3)$ & $0.3276(7)$ \\
    \hline
    s1-sp3 & $20$ & $17900$ & $2$ & $0.718679(3)$ & $0.600098(4)$ & $0.20501(11)$ & $0.47(5)$ & $0.248(3)$ & $0.3265(3)$ \\
    \hline
    s1-sp3 & $24$ & $21802$ & $2$ & $0.718678(2)$ & $0.600095(3)$ & $0.20433(9)$ & $0.45(6)$ & $0.244(3)$ & $0.3258(3)$ \\
    \hline
    s1-sp3 & $28$ & $10743$ & $2$ & $0.718676(2)$ & $0.600098(3)$ & $0.20397(6)$ & $0.44(9)$ & $0.235(4)$ & $0.3260(4)$ \\
    \hline\hline
    s1-sp4 & $12$ & $19860$ & $4$ & $0.719618(8)$ & $0.600155(10)$ & $0.2089(4)$ & $0.48(3)$ & $0.264(4)$ & $0.2479(5)$ \\
    \hline
    s1-sp4 & $16$ & $17885$ & $4$ & $0.719668(5)$ & $0.600250(6)$ & $0.2070(3)$ & $0.47(3)$ & $0.251(3)$ & $0.2335(4)$ \\
    \hline
    s1-sp4 & $20$ & $17900$ & $2$ & $0.719686(3)$ & $0.600279(4)$ & $0.20668(11)$ & $0.47(5)$ & $0.248(2)$ & $0.2301(4)$ \\
    \hline
    s1-sp4 & $24$ & $19918$ & $2$ & $0.719687(2)$ & $0.600273(3)$ & $0.20611(9)$ & $0.50(4)$ & $0.251(2)$ & $0.22925(14)$ \\
    \hline
    s1-sp4 & $28$ & $9867$ & $2$ & $0.719687(2)$ & $0.600277(3)$ & $0.20580(6)$ & $0.48(9)$ & $0.246(4)$ & $0.2284(5)$ \\
    \hline\hline
    s1-sp5 & $12$ & $19860$ & $8$ & $0.721129(8)$ & $0.600280(11)$ & $0.2097(5)$ & $0.387(12)$ & $0.214(3)$ & $0.1320(7)$ \\
    \hline
    s1-sp5 & $16$ & $14885$ & $8$ & $0.721267(5)$ & $0.600505(7)$ & $0.2093(3)$ & $0.44(3)$ & $0.212(3)$ & $0.1042(5)$ \\
    \hline
    s1-sp5 & $20$ & $11900$ & $4$ & $0.721314(4)$ & $0.600552(5)$ & $0.20932(13)$ & $0.45(3)$ & $0.223(3)$ & $0.0937(2)$ \\
    \hline
    s1-sp5 & $24$ & $19754$ & $4$ & $0.721333(2)$ & $0.600570(3)$ & $0.20902(9)$ & $0.40(5)$ & $0.224(3)$ & $0.09015(14)$ \\
    \hline
    s1-sp5 & $28$ & $8038$ & $2$ & $0.721343(2)$ & $0.600579(3)$ & $0.20882(6)$ & $0.53(4)$ & $0.227(4)$ & $0.0886(5)$ \\
    \hline
\end{tabular}
\caption{Lattice data for all ensembles of set~1. All quantities are in lattice units. Only statistical errors are quoted.}
\label{tab:data_s1}
\end{table}

\begin{table}
\centering
\begin{tabular}{| c | c | c | c || c | c | c || c | c | c |}
    \hline
    Ensemble & $L$ & $N_{\textrm{cnfg}}$ & $N_{\textrm{bin}}$ & $\langle P \rangle$ & $\langle \sqrt{\phi^\dagger \phi} \rangle$ & $a v$ &$am_{0^+}$ & $am_{1^-}$ & $am_{\Psi}$ \\
    \hline\hline
    s2-sp1 & $12$ & $19860$ & $2$ & $0.802968(5)$ & $0.647031(10)$ & $0.36666(14)$ & $0.58(2)$ & $0.328(3)$ & $1.3884(8)$ \\
    \hline
    s2-sp1 & $16$ & $18885$ & $2$ & $0.802964(3)$ & $0.647039(6)$ & $0.36583(7)$ & $0.62(3)$ & $0.324(3)$ & $1.3871(9)$ \\
    \hline
    s2-sp1 & $20$ & $19925$ & $2$ & $0.802969(2)$ & $0.647039(4)$ & $0.36535(5)$ & $0.640(11)$ & $0.317(3)$ & $1.3870(8)$ \\
    \hline
    s2-sp1 & $24$ & $21918$ & $2$ & $0.8029692(12)$ & $0.647039(3)$ & $0.36513(3)$ & $0.674(7)$ & $0.320(3)$ & $1.3870(8)$ \\
    \hline
    s2-sp1 & $28$ & $9915$ & $2$ & $0.8029678(12)$ & $0.647038(3)$ & $0.36498(3)$ & $0.660(9)$ & $0.321(3)$ & $1.3886(12)$ \\
    \hline\hline
    s2-sp2 & $12$ & $19860$ & $2$ & $0.803655(5)$ & $0.647117(10)$ & $0.36710(13)$ & $0.63(2)$ & $0.329(2)$ & $0.7088(3)$ \\
    \hline
    s2-sp2 & $16$ & $19885$ & $2$ & $0.803652(3)$ & $0.647127(5)$ & $0.36616(8)$ & $0.61(2)$ & $0.319(3)$ & $0.7069(3)$ \\
    \hline
    s2-sp2 & $20$ & $19900$ & $2$ & $0.803651(2)$ & $0.647127(4)$ & $0.36571(5)$ & $0.65(4)$ & $0.314(2)$ & $0.7062(2)$ \\
    \hline
    s2-sp2 & $24$ & $19910$ & $2$ & $0.8036524(12)$ & $0.647128(3)$ & $0.36544(4)$ & $0.664(7)$ & $0.319(4)$ & $0.7067(2)$ \\
    \hline
    s2-sp2 & $28$ & $11186$ & $2$ & $0.8036543(12)$ & $0.647127(3)$ & $0.36529(3)$ & $0.68(4)$ & $0.321(4)$ & $0.7063(2)$ \\
    \hline\hline
    s2-sp3 & $12$ & $19860$ & $2$ & $0.804095(5)$ & $0.647163(9)$ & $0.36736(13)$ & $0.56(2)$ & $0.328(2)$ & $0.5106(4)$ \\
    \hline
    s2-sp3 & $16$ & $19885$ & $2$ & $0.804090(3)$ & $0.647187(6)$ & $0.36639(8)$ & $0.63(2)$ & $0.321(3)$ & $0.50927(14)$ \\
    \hline
    s2-sp3 & $20$ & $20481$ & $2$ & $0.804093(2)$ & $0.647187(3)$ & $0.36598(5)$ & $0.65(2)$ & $0.321(3)$ & $0.50914(12)$ \\
    \hline
    s2-sp3 & $24$ & $20589$ & $2$ & $0.8040918(12)$ & $0.647187(3)$ & $0.36576(4)$ & $0.641(12)$ & $0.316(4)$ & $0.50906(10)$ \\
    \hline
    s2-sp3 & $28$ & $10933$ & $2$ & $0.8040914(11)$ & $0.647186(3)$ & $0.36556(3)$ & $0.64(3)$ & $0.319(4)$ & $0.50887(11)$ \\
    \hline\hline
    s2-sp4 & $12$ & $19860$ & $2$ & $0.804522(5)$ & $0.647221(9)$ & $0.36749(13)$ & $0.602(10)$ & $0.318(4)$ & $0.36371(14)$ \\
    \hline
    s2-sp4 & $16$ & $17885$ & $2$ & $0.804529(3)$ & $0.647243(6)$ & $0.36666(8)$ & $0.61(3)$ & $0.318(3)$ & $0.36153(10)$ \\
    \hline
    s2-sp4 & $20$ & $19900$ & $2$ & $0.804532(2)$ & $0.647248(4)$ & $0.36620(5)$ & $0.63(2)$ & $0.318(3)$ & $0.36094(11)$ \\
    \hline
    s2-sp4 & $24$ & $19920$ & $2$ & $0.8045299(12)$ & $0.647254(3)$ & $0.36600(4)$ & $0.658(9)$ & $0.321(3)$ & $0.36100(6)$ \\
    \hline
    s2-sp4 & $28$ & $10482$ & $2$ & $0.8045320(12)$ & $0.647249(3)$ & $0.36581(3)$ & $0.67(2)$ & $0.311(4)$ & $0.36112(11)$ \\
    \hline\hline
    s2-sp5 & $12$ & $19860$ & $4$ & $0.805441(5)$ & $0.647335(10)$ & $0.36801(12)$ & $0.553(9)$ & $0.289(3)$ & $0.1271(2)$ \\
    \hline
    s2-sp5 & $16$ & $14885$ & $2$ & $0.805480(4)$ & $0.647382(6)$ & $0.36723(8)$ & $0.620(10)$ & $0.294(4)$ & $0.11932(12)$ \\
    \hline
    s2-sp5 & $20$ & $19900$ & $2$ & $0.805501(2)$ & $0.647386(4)$ & $0.36680(5)$ & $0.60(2)$ & $0.302(3)$ & $0.11697(5)$ \\
    \hline
    s2-sp5 & $24$ & $21799$ & $2$ & $0.8055021(11)$ & $0.647392(3)$ & $0.36660(4)$ & $0.66(2)$ & $0.307(4)$ & $0.11631(5)$ \\
    \hline
    s2-sp5 & $28$ & $9935$ & $2$ & $0.8055050(12)$ & $0.647391(3)$ & $0.36641(3)$ & $0.65(2)$ & $0.319(4)$ & $0.11593(6)$ \\
    \hline
\end{tabular}
\caption{Lattice data for all ensembles of set~2. All quantities are in lattice units. Only statistical errors are quoted.}
\label{tab:data_s2}
\end{table}

\begin{table}
\centering
\begin{tabular}{| c | c | c | c || c | c | c || c | c | c |}
    \hline
    Ensemble & $L$ & $N_{\textrm{cnfg}}$ & $N_{\textrm{bin}}$ & $\langle P \rangle$ & $\langle \sqrt{\phi^\dagger \phi} \rangle$ & $a v$ &$am_{0^+}$ & $am_{1^-}$ & $am_{\Psi}$ \\
    \hline\hline
    s3-sp1 & $12$ & $19800$ & $8$ & $0.904478(3)$ & $0.63276(13)$ & $0.3603(12)$ & $0.293(4)$ & $0.211(2)$ & $0.3873(3)$ \\
    \hline
    s3-sp1 & $16$ & $18885$ & $6$ & $0.904487(2)$ & $0.63415(7)$ & $0.3634(6)$ & $0.284(12)$ & $0.2023(12)$ & $0.38462(13)$ \\
    \hline
    s3-sp1 & $20$ & $19900$ & $6$ & $0.9044887(9)$ & $0.63450(4)$ & $0.3624(4)$ & $0.295(7)$ & $0.1983(12)$ & $0.38348(10)$ \\
    \hline
    s3-sp1 & $24$ & $17910$ & $4$ & $0.9044899(6)$ & $0.63459(3)$ & $0.3626(3)$ & $0.288(11)$ & $0.1975(13)$ & $0.38338(10)$ \\
    \hline
    s3-sp1 & $28$ & $9992$ & $4$ & $0.9044880(7)$ & $0.63453(3)$ & $0.3623(2)$ & $0.316(6)$ & $0.196(2)$ & $0.3826(3)$ \\
    \hline
    s3-sp1 & $32$ & $4295$ & $4$ & $0.9044887(8)$ & $0.63456(3)$ &   & $0.317(9)$ & $0.200(3)$ & $0.3832(4)$ \\
    \hline\hline
    s3-sp2 & $12$ & $19000$ & $8$ & $0.904624(3)$ & $0.63347(14)$ & $0.3624(12)$ & $0.304(5)$ & $0.201(3)$ & $0.1937(3)$ \\
    \hline
    s3-sp2 & $16$ & $19885$ & $6$ & $0.904631(2)$ & $0.63492(7)$ & $0.3649(6)$ & $0.290(11)$ & $0.2003(12)$ & $0.18832(14)$ \\
    \hline
    s3-sp2 & $20$ & $20000$ & $6$ & $0.9046345(9)$ & $0.63522(4)$ & $0.3643(4)$ & $0.316(5)$ & $0.1984(12)$ & $0.18705(8)$ \\
    \hline
    s3-sp2 & $24$ & $17910$ & $4$ & $0.9046369(7)$ & $0.63532(3)$ & $0.3646(3)$ & $0.315(6)$ & $0.1976(13)$ & $0.18637(6)$ \\
    \hline
    s3-sp2 & $28$ & $7916$ & $4$ & $0.9046367(7)$ & $0.63541(3)$ & $0.36432(13)$ & $0.308(7)$ & $0.200(2)$ & $0.18609(13)$ \\
    \hline
    s3-sp2 & $32$ & $3420$ & $4$ & $0.9046359(8)$ & $0.63531(3)$ &   & $0.298(10)$ & $0.190(4)$ & $0.1863(3)$ \\
    \hline\hline
    s3-sp3 & $12$ & $19700$ & $8$ & $0.904671(3)$ & $0.63324(14)$ & $0.3589(12)$ & $0.303(5)$ & $0.196(2)$ & $0.1335(3)$ \\
    \hline
    s3-sp3 & $16$ & $19235$ & $6$ & $0.904687(2)$ & $0.63509(7)$ & $0.3645(5)$ & $0.292(5)$ & $0.1964(12)$ & $0.12504(11)$ \\
    \hline
    s3-sp3 & $20$ & $19900$ & $6$ & $0.9046924(9)$ & $0.63556(4)$ & $0.3650(4)$ & $0.301(5)$ & $0.1966(13)$ & $0.12267(9)$ \\
    \hline
    s3-sp3 & $24$ & $17022$ & $4$ & $0.9046930(7)$ & $0.63565(3)$ & $0.3651(3)$ & $0.297(6)$ & $0.1966(13)$ & $0.12189(5)$ \\
    \hline
    s3-sp3 & $28$ & $7102$ & $4$ & $0.9046938(7)$ & $0.63566(3)$ & $0.36486(14)$ & $0.305(7)$ & $0.195(3)$ & $0.12156(7)$ \\
    \hline
    s3-sp3 & $32$ & $2740$ & $4$ & $0.9046925(9)$ & $0.63556(4)$ &   & $0.319(9)$ & $0.193(3)$ & $0.12149(7)$ \\
    \hline\hline
    s3-sp4 & $12$ & $18000$ & $8$ & $0.904680(3)$ & $0.63327(14)$ & $0.3609(11)$ & $0.307(5)$ & $0.193(2)$ & $0.1190(3)$ \\
    \hline
    s3-sp4 & $16$ & $17885$ & $6$ & $0.904699(2)$ & $0.63524(7)$ & $0.3645(6)$ & $0.298(9)$ & $0.1940(12)$ & $0.1106(3)$ \\
    \hline
    s3-sp4 & $20$ & $20932$ & $6$ & $0.9047044(9)$ & $0.63551(4)$ & $0.3655(4)$ & $0.293(12)$ & $0.1943(12)$ & $0.10791(9)$ \\
    \hline
    s3-sp4 & $24$ & $15199$ & $4$ & $0.9047078(7)$ & $0.63570(3)$ & $0.3650(3)$ & $0.308(6)$ & $0.1955(14)$ & $0.1065(2)$ \\
    \hline
    s3-sp4 & $28$ & $5346$ & $4$ & $0.9047085(8)$ & $0.63579(4)$ & $0.36522(14)$ & $0.322(9)$ & $0.201(3)$ & $0.10660(7)$ \\
    \hline
    s3-sp4 & $32$ & $1927$ & $4$ & $0.9047073(11)$ & $0.63572(4)$ &   & $0.310(12)$ & $0.200(5)$ & $0.10641(8)$ \\
    \hline\hline
    s3-sp5 & $12$ & $17000$ & $8$ & $0.904701(3)$ & $0.6324(2)$ & $0.3595(11)$ & $0.296(5)$ & $0.177(2)$ & $0.090(2)$ \\
    \hline
    s3-sp5 & $16$ & $14885$ & $6$ & $0.904733(2)$ & $0.63516(8)$ & $0.3652(6)$ & $0.289(7)$ & $0.1842(13)$ & $0.0766(3)$ \\
    \hline
    s3-sp5 & $20$ & $21159$ & $6$ & $0.9047403(9)$ & $0.63570(4)$ & $0.3653(4)$ & $0.301(5)$ & $0.1901(12)$ & $0.07244(10)$ \\
    \hline
    s3-sp5 & $24$ & $10576$ & $4$ & $0.9047431(8)$ & $0.63597(4)$ & $0.3656(2)$ & $0.293(10)$ & $0.192(2)$ & $0.07097(7)$ \\
    \hline
    s3-sp5 & $28$ & $3238$ & $4$ & $0.9047435(11)$ & $0.63598(5)$ & $0.36543(14)$ & $0.319(10)$ & $0.190(3)$ & $0.07032(7)$ \\
    \hline
    s3-sp5 & $32$ & $1519$ & $4$ & $0.9047436(12)$ & $0.63593(5)$ &   & $0.317(12)$ & $0.195(6)$ & $0.07002(8)$ \\
    \hline
\end{tabular}
\caption{Lattice data for all ensembles of set~3. All quantities are in lattice units. Only statistical errors are quoted.}
\label{tab:data_s3}
\end{table}

\bibliographystyle{apsrev4-1}
\bibliography{main.bib}

@article{data_release,
  author       = {Maas, Axel and Martins, Sofie and Wieland, Georg},
  title        = {{Spectroscopy of the SU(2) scalar-fermion-gauge system under the influence of the Brout-Englert-Higgs effect -- data release}},
  journal        = {{Spectroscopy of the SU(2) scalar-fermion-gauge system under the influence of the Brout-Englert-Higgs effect -- data release}},
  month        = oct,
  year         = 2026,
  publisher    = {Zenodo},
  version      = {1.0.0},
  doi          = {10.5281/zenodo.23319057},
  url          = {https://doi.org/10.5281/zenodo.23319057}
}

@article{Cabibbo:1982zn,
    author = "Cabibbo, N. and Marinari, E.",
    title = "{A New Method for Updating SU(N) Matrices in Computer Simulations of Gauge Theories}",
    doi = "10.1016/0370-2693(82)90696-7",
    journal = "Phys. Lett. B",
    volume = "119",
    pages = "387--390",
    year = "1982"
}

@article{Maas:2017wzi,
    author = "Maas, Axel",
    title = "{Brout-Englert-Higgs physics: From foundations to phenomenology}",
    eprint = "1712.04721",
    archivePrefix = "arXiv",
    primaryClass = "hep-ph",
    doi = "10.1016/j.ppnp.2019.02.003",
    journal = "Prog. Part. Nucl. Phys.",
    volume = "106",
    pages = "132--209",
    year = "2019"
}

@article{Frohlich:1980gj,
    author = "Fr{\"o}hlich, J. and Morchio, G. and Strocchi, F.",
    title = "{Higgs phenomenon without a symmetry breaking order parameter}",
    reportNumber = "Print-80-0557 (SNS,PISA)",
    doi = "10.1016/0370-2693(80)90594-8",
    journal = "Phys. Lett. B",
    volume = "97",
    pages = "249--252",
    year = "1980"
}

@article{Frohlich:1981yi,
    author = "Fr{\"o}hlich, J. and Morchio, G. and Strocchi, F.",
    title = "{Higgs phenomenon without symmetry breaking order parameter}",
    doi = "10.1016/0550-3213(81)90448-X",
    journal = "Nucl. Phys. B",
    volume = "190",
    pages = "553--582",
    year = "1981"
}

@article{Dudal:2020uwb,
    author = "Dudal, D. and van Egmond, D. M. and Guimaraes, M. S. and Palhares, L. F. and Peruzzo, G. and Sorella, S. P.",
    title = "{Spectral properties of local gauge invariant composite operators in the $SU(2)$ Yang{\textendash}Mills{\textendash}Higgs model}",
    eprint = "2008.07813",
    archivePrefix = "arXiv",
    primaryClass = "hep-th",
    doi = "10.1140/epjc/s10052-021-09008-9",
    journal = "Eur. Phys. J. C",
    volume = "81",
    number = "3",
    pages = "222",
    year = "2021"
}

@article{Maas:2013aia,
    author = "Maas, Axel and Mufti, Tajdar",
    title = "{Two- and three-point functions in Landau gauge Yang-Mills-Higgs theory}",
    eprint = "1312.4873",
    archivePrefix = "arXiv",
    primaryClass = "hep-lat",
    doi = "10.1007/JHEP04(2014)006",
    journal = "JHEP",
    volume = "04",
    pages = "006",
    year = "2014"
}

@article{Afferrante:2020fhd,
    author = {Afferrante, Vincenzo and Maas, Axel and Sondenheimer, Ren{\'e} and T{\"o}rek, Pascal},
    title = "{Testing the mechanism of lepton compositeness}",
    eprint = "2011.02301",
    archivePrefix = "arXiv",
    primaryClass = "hep-lat",
    doi = "10.21468/SciPostPhys.10.3.062",
    journal = "SciPost Phys.",
    volume = "10",
    number = "3",
    pages = "062",
    year = "2021"
}

@article{DelDebbio:2008zf,
    author = "Del Debbio, Luigi and Patella, Agostino and Pica, Claudio",
    title = "{Higher representations on the lattice: Numerical simulations. SU(2) with adjoint fermions}",
    eprint = "0805.2058",
    archivePrefix = "arXiv",
    primaryClass = "hep-lat",
    reportNumber = "NI08019, BNL-NT-08-15, CP3-ORIGINS-2010-15",
    doi = "10.1103/PhysRevD.81.094503",
    journal = "Phys. Rev. D",
    volume = "81",
    pages = "094503",
    year = "2010"
}

@article{Gattringer:2008je,
    author = "Gattringer, Christof and Pak, Markus",
    title = "{New overlap construction of Weyl fermions on the lattice}",
    eprint = "0802.2496",
    archivePrefix = "arXiv",
    primaryClass = "hep-lat",
    doi = "10.1016/j.nuclphysb.2008.03.015",
    journal = "Nucl. Phys. B",
    volume = "801",
    pages = "353--360",
    year = "2008"
}

@Article{Greensite:2020lmh,
  author        = {Greensite, Jeff},
  journal       = {Phys. Rev. D},
  title         = {{Excitations of elementary fermions in gauge Higgs theories}},
  year          = {2020},
  number        = {5},
  pages         = {054504},
  volume        = {102},
  archiveprefix = {arXiv},
  doi           = {10.1103/PhysRevD.102.054504},
  eprint        = {2007.11616},
  primaryclass  = {hep-lat},
}

@article{Wieland:2025tog,
    author = "Wieland, Georg and Maas, Axel",
    title = "{Spectrum of the SU(2) scalar-fermion-gauge system under the influence of the Brout-Englert-Higgs effect}",
    eprint = "2510.27457",
    archivePrefix = "arXiv",
    primaryClass = "hep-lat",
    doi = "10.22323/1.485.0296",
    journal = "PoS",
    volume = "EPS-HEP2025",
    pages = "296",
    year = "2026"
}

@BOOK{Shifman:2012zz,
  title = {{Advanced topics in quantum field theory: A lecture course}},
  publisher = {Cambridge University Press},
  year = {2012},
  author = {Shifman, M.},
  pages = {622},
  slaccitation = {%%CITATION = INSPIRE-1115926;%%}
}

@inproceedings{Martins:2026eni,
    author = "Martins, Sofie and Jenny, Patrick and Maas, Axel and Wieland, Georg",
    title = "{Weak and Higgs physics from the lattice}",
    booktitle = "{42th International Symposium on Lattice Field Theory}",
    eprint = "2603.12882",
    archivePrefix = "arXiv",
    primaryClass = "hep-lat",
    month = "3",
    year = "2026"
}

@book{Bambi:2015mba,
    author = "Bambi, Cosimo and Dolgov, Alexandre D.",
    title = "{Introduction to Particle Cosmology. The Standard Model of Cosmology and its Open Problems}",
    doi = "10.1007/978-3-662-48078-6",
    isbn = "978-3-662-48077-9, 978-3-662-51587-7, 978-3-662-48078-6",
    publisher = "Springer",
    series = "UNITEXT for Physics",
    year = "2016"
}

@article{Maas:2018xxu,
    author = {Maas, Axel and T{\"o}rek, Pascal},
    title = "{The spectrum of an SU(3) gauge theory with a fundamental Higgs field}",
    eprint = "1804.04453",
    archivePrefix = "arXiv",
    primaryClass = "hep-lat",
    doi = "10.1016/j.aop.2018.08.018",
    journal = "Annals Phys.",
    volume = "397",
    pages = "303--335",
    year = "2018"
}

@article{Maas:2012ct,
    author = "Maas, Axel",
    title = "{(Non-)Aligned gauges and global gauge symmetry breaking}",
    eprint = "1205.0890",
    archivePrefix = "arXiv",
    primaryClass = "hep-th",
    doi = "10.1142/S0217732312502227",
    journal = "Mod. Phys. Lett. A",
    volume = "27",
    pages = "1250222",
    year = "2012"
}

@book{Kapusta:2006pm,
    author = "Kapusta, J. I. and Gale, Charles",
    title = "{Finite-temperature field theory: Principles and applications}",
    doi = "10.1017/CBO9780511535130",
    isbn = "978-0-521-17322-3, 978-0-521-82082-0, 978-0-511-22280-1",
    publisher = "Cambridge University Press",
    series = "Cambridge Monographs on Mathematical Physics",
    year = "2011"
}

@article{Drach:2017btk,
    author = "Drach, Vincent and Janowski, Tadeusz and Pica, Claudio",
    editor = "Della Morte, M. and Fritzsch, P. and G{\'a}miz S{\'a}nchez, E. and Pena Ruano, C.",
    title = "{Update on SU(2) gauge theory with NF = 2 fundamental flavours}",
    eprint = "1710.07218",
    archivePrefix = "arXiv",
    primaryClass = "hep-lat",
    doi = "10.1051/epjconf/201817508020",
    journal = "EPJ Web Conf.",
    volume = "175",
    pages = "08020",
    year = "2018"
}

@article{Caudy:2007sf,
    author = "Caudy, W. and Greensite, J.",
    title = "{On the ambiguity of spontaneously broken gauge symmetry}",
    eprint = "0712.0999",
    archivePrefix = "arXiv",
    primaryClass = "hep-lat",
    doi = "10.1103/PhysRevD.78.025018",
    journal = "Phys. Rev. D",
    volume = "78",
    pages = "025018",
    year = "2008"
}

@article{Afferrante:2020hqe,
    author = {Afferrante, Vincenzo and Maas, Axel and T{\"o}rek, Pascal},
    title = "{Composite massless vector boson}",
    eprint = "2002.08221",
    archivePrefix = "arXiv",
    primaryClass = "hep-lat",
    doi = "10.1103/PhysRevD.101.114506",
    journal = "Phys. Rev. D",
    volume = "101",
    number = "11",
    pages = "114506",
    year = "2020"
}

@article{Greensite:2026jim,
    author = "Greensite, Jeff",
    title = "{Varieties of electrically charged physical states in SU(2)$\times$U(1) lattice gauge Higgs theory}",
    eprint = "2604.01166",
    archivePrefix = "arXiv",
    primaryClass = "hep-lat",
    month = "4",
    year = "2026"
}

@article{vonSmekal:2012vx,
    author = "von Smekal, L.",
    editor = "Schaefer, B. -J. and Alkofer, R. and Fischer, C. S.",
    title = "{Universal Aspects of QCD-like Theories}",
    eprint = "1205.4205",
    archivePrefix = "arXiv",
    primaryClass = "hep-ph",
    doi = "10.1016/j.nuclphysbps.2012.06.006",
    journal = "Nucl. Phys. B Proc. Suppl.",
    volume = "228",
    pages = "179--220",
    year = "2012"
}

@Article{Andersen:2011yj,
  author        = {Andersen, J.R. and Antipin, O. and Azuelos, G. and Del Debbio, L. and Del Nobile, E. and others},
  journal       = {Eur.Phys.J.Plus},
  title         = {{Discovering Technicolor}},
  year          = {2011},
  pages         = {81},
  volume        = {126},
  archiveprefix = {arXiv},
  doi           = {10.1140/epjp/i2011-11081-1},
  eprint        = {1104.1255},
  primaryclass  = {hep-ph},
}

@article{pdg,
    author = "Takahashi, F. and others",
    collaboration = "Particle Data Group",
    title = "{Review of Particle Physics}",
    doi = "10.1142/S0217751X26300115",
    journal = "Int. J. Mod. Phys. A",
    volume = "41",
    pages = "2630011",
    year = "2026"
}

@Article{Hasenfratz:2007dp,
  author =        {Hasenfratz, Peter and von Allmen, Reto},
  title =         {{Towards Weyl fermions on the lattice without artefacts}},
  journal =       {JHEP},
  year =          {2008},
  volume =        {02},
  pages =         {079},
  archiveprefix = {arXiv},
  doi =           {10.1088/1126-6708/2008/02/079},
  eprint =        {0710.5346},
  primaryclass =  {hep-lat},
  slaccitation =  {%%CITATION = ARXIV:0710.5346;%%}
}

@Article{Hansen:2017mrt,
  author =        {Hansen, Martin and Janowski, Tadeusz and Pica, Claudio and Toniato, Arianna},
  title =         {{SU(2) with fundamental fermions and scalars}},
  journal =       {EPJ Web Conf.},
  year =          {2018},
  volume =        {175},
  pages =         {08010},
  archiveprefix = {arXiv},
  booktitle =     {{Proceedings, 35th International Symposium on Lattice Field Theory (Lattice 2017): Granada, Spain, June 18-24, 2017}},
  doi =           {10.1051/epjconf/201817508010},
  eprint =        {1710.10831},
  primaryclass =  {hep-lat},
  slaccitation =  {%%CITATION = ARXIV:1710.10831;%%}
}

@article{Greensite:2025fdq,
    author = "Greensite, Jeff",
    title = "{Does the Z boson have a lighter cousin?}",
    eprint = "2501.02083",
    archivePrefix = "arXiv",
    primaryClass = "hep-lat",
    doi = "10.1103/PhysRevD.111.073005",
    journal = "Phys. Rev. D",
    volume = "111",
    number = "7",
    pages = "073005",
    year = "2025"
}

@article{Sondenheimer:2019idq,
    author = "Sondenheimer, Ren{\'e}",
    title = "{Analytical relations for the bound state spectrum of gauge theories with a Brout-Englert-Higgs mechanism}",
    eprint = "1912.08680",
    archivePrefix = "arXiv",
    primaryClass = "hep-th",
    doi = "10.1103/PhysRevD.101.056006",
    journal = "Phys. Rev. D",
    volume = "101",
    number = "5",
    pages = "056006",
    year = "2020"
}

@article{deBlas:2025gyz,
    author = "de Blas, Jorge and others",
    title = "{Physics Briefing Book: Input for the 2026 update of the European Strategy for Particle Physics}",
    eprint = "2511.03883",
    archivePrefix = "arXiv",
    primaryClass = "hep-ex",
    reportNumber = "CERN--2025-008, CERN-ESU-2025-001",
    doi = "10.23731/CYRM-2025-008",
    month = "11",
    year = "2025"
}

@Article{Lee:1987eg,
  author       = {Lee, I-Hsiu and Shrock, Robert E.},
  journal      = {Phys. Rev. Lett.},
  title        = {{Chiral Symmetry Breaking Phase Transition in Lattice Gauge Higgs Theories With Fermions}},
  year         = {1987},
  pages        = {14},
  volume       = {59},
  doi          = {10.1103/PhysRevLett.59.14},
  reportnumber = {ITP-SB-87-16, BNL-39565},
}

@Article{Lee:1988ut,
  author =       {Lee, I-Hsiu and Shrock, Robert E.},
  title =        {{The Chiral Transition in an SU(2) Gauge Higgs Fermion Theory With Yukawa Couplings}},
  journal =      {Nucl. Phys.},
  year =         {1988},
  volume =       {B305},
  pages =        {305-323},
  doi =          {10.1016/0550-3213(88)90298-2},
  reportnumber = {ITP-SB-88-27},
  slaccitation = {%%CITATION = NUPHA,B305,305;%%}
}

@inbook{Maas:2023emb,
    author = "Maas, Axel",
    editor = "Cintio, Andrea and Michelangeli, Alessandro",
    title = {{The Fr{\"o}hlich-Morchio-Strocchi mechanism: an underestimated legacy}},
    booktitle = "{Trails in Modern Theoretical and Mathematical Physics}",
    eprint = "2305.01960",
    archivePrefix = "arXiv",
    primaryClass = "hep-th",
    doi = "10.1007/978-3-031-44988-8_11",
    pages = "177--205",
    month = "5",
    year = "2023"
}

@article{Maas:2024hcn,
    author = {Maas, Axel and van Egmond, Duifje M. and Pl{\"a}tzer, Simon},
    title = "{Subleading Higgs effects at lepton colliders}",
    eprint = "2409.20131",
    archivePrefix = "arXiv",
    primaryClass = "hep-ph",
    doi = "10.22323/1.476.0056",
    journal = "PoS",
    volume = "ICHEP2024",
    pages = "056",
    year = "2025"
}

@article{Jenny:2022atm,
    author = "Jenny, Patrick and Maas, Axel and Riederer, Bernd",
    title = "{Vector boson scattering from the lattice}",
    eprint = "2204.02756",
    archivePrefix = "arXiv",
    primaryClass = "hep-lat",
    doi = "10.1103/PhysRevD.105.114513",
    journal = "Phys. Rev. D",
    volume = "105",
    number = "11",
    pages = "114513",
    year = "2022"
}

@article{Maas:2020kda,
    author = "Maas, Axel and Sondenheimer, Ren{\'e}",
    title = "{Gauge-invariant description of the Higgs resonance and its phenomenological implications}",
    eprint = "2009.06671",
    archivePrefix = "arXiv",
    primaryClass = "hep-ph",
    doi = "10.1103/PhysRevD.102.113001",
    journal = "Phys. Rev. D",
    volume = "102",
    pages = "113001",
    year = "2020"
}

@article{Egger:2017tkd,
    author = "Egger, Larissa and Maas, Axel and Sondenheimer, Ren{\'e}",
    title = "{Pair production processes and flavor in gauge-invariant perturbation theory}",
    eprint = "1701.02881",
    archivePrefix = "arXiv",
    primaryClass = "hep-ph",
    doi = "10.1142/S0217732317502121",
    journal = "Mod. Phys. Lett. A",
    volume = "32",
    number = "38",
    pages = "1750212",
    year = "2017"
}

@article{Maas:2022gdb,
    author = "Maas, Axel and Reiner, Franziska",
    title = "{Restoring the Bloch-Nordsieck theorem in the electroweak sector of the standard model}",
    eprint = "2212.08470",
    archivePrefix = "arXiv",
    primaryClass = "hep-ph",
    doi = "10.1103/PhysRevD.108.013001",
    journal = "Phys. Rev. D",
    volume = "108",
    number = "1",
    pages = "013001",
    year = "2023"
}

@article{Maas:2012tj,
    author = "Maas, Axel",
    title = "{Bound-state/elementary-particle duality in the Higgs sector and the case for an excited 'Higgs' within the standard model}",
    eprint = "1205.6625",
    archivePrefix = "arXiv",
    primaryClass = "hep-lat",
    doi = "10.1142/S0217732313501034",
    journal = "Mod. Phys. Lett. A",
    volume = "28",
    pages = "1350103",
    year = "2013"
}

@article{Lee:1987zu,
    author = "Lee, I-Hsiu and Shrock, Robert E.",
    title = "{Analytic and Numerical Results on the Chiral Transition in SU(2) Lattice Gauge Higgs Theory With Fermions}",
    reportNumber = "ITP-SB-87-68, BNL-40515",
    doi = "10.1016/0370-2693(88)90607-7",
    journal = "Phys. Lett. B",
    volume = "201",
    pages = "497--502",
    year = "1988"
}

@article{Aoki:1988fg,
    author = "Aoki, Sinya and Lee, I-Hsiu and Shrock, Robert E.",
    title = "{SU(2) lattice gauge-Higgs theory with dynamical I = 1 fermions and the role of Fermion representation in chiral phase structure}",
    reportNumber = "ITP-SB-88-46",
    doi = "10.1016/0370-2693(89)90400-0",
    journal = "Phys. Lett. B",
    volume = "219",
    pages = "335--341",
    year = "1989"
}

@article{Elitzur:1975im,
    author = "Elitzur, S.",
    title = "{Impossibility of Spontaneously Breaking Local Symmetries}",
    doi = "10.1103/PhysRevD.12.3978",
    journal = "Phys. Rev. D",
    volume = "12",
    pages = "3978--3982",
    year = "1975"
}

@book{Bohm:2001yx,
    author = "B{\"o}hm, M. and Denner, Ansgar and Joos, H.",
    title = "{Gauge theories of the strong and electroweak interaction}",
    doi = "10.1007/978-3-322-80160-9",
    isbn = "978-3-519-23045-8, 978-3-322-80162-3, 978-3-322-80160-9",
    publisher = "Teubner",
    address = "Stuttgart",
    year = "2001"
}

@book{Montvay:1994cy,
    author = "Montvay, I. and M{\"u}nster, G.",
    title = "{Quantum fields on a lattice}",
    doi = "10.1017/CBO9780511470783",
    isbn = "978-0-521-59917-7, 978-0-511-87919-7",
    publisher = "Cambridge University Press",
    address = "Cambridge",
    year = "1997"
}

@book{Gattringer:2010zz,
    author = "Gattringer, Christof and Lang, Christian B.",
    title = "{Quantum chromodynamics on the lattice}",
    doi = "10.1007/978-3-642-01850-3",
    isbn = "978-3-642-01849-7, 978-3-642-01850-3",
    publisher = "Springer",
    address = "Berlin",
    year = "2010"
}

@article{Evertz:1985fc,
    author = "Evertz, H. G. and Jersak, J. and Lang, C. B. and Neuhaus, T.",
    title = "{SU(2) Higgs Boson and Vector Boson Masses on the Lattice}",
    reportNumber = "PITHA 85/23",
    doi = "10.1016/0370-2693(86)91547-9",
    journal = "Phys. Lett. B",
    volume = "171",
    pages = "271--279",
    year = "1986"
}

@article{Wurtz:2013ova,
    author = "Wurtz, Mark and Lewis, Randy",
    title = "{Higgs and W boson spectrum from lattice simulations}",
    eprint = "1307.1492",
    archivePrefix = "arXiv",
    primaryClass = "hep-lat",
    doi = "10.1103/PhysRevD.88.054510",
    journal = "Phys. Rev. D",
    volume = "88",
    pages = "054510",
    year = "2013"
}

@article{Drach:2025eyg,
    author = "Drach, Vincent and Martins, Sofie and Pica, Claudio and Rago, Antonio",
    title = "{High-performance simulations of higher representations of Wilson fermions}",
    eprint = "2503.06721",
    archivePrefix = "arXiv",
    primaryClass = "hep-lat",
    doi = "10.1016/j.cpc.2026.110061",
    journal = "Comput. Phys. Commun.",
    volume = "322",
    pages = "110061",
    year = "2026"
}

@article{Luscher:1985dn,
    author = "L{\"u}scher, M.",
    title = "{Volume Dependence of the Energy Spectrum in Massive Quantum Field Theories. 1. Stable Particle States}",
    reportNumber = "DESY-85-144",
    doi = "10.1007/BF01211589",
    journal = "Commun. Math. Phys.",
    volume = "104",
    pages = "177",
    year = "1986"
}

@article{Luscher:1986pf,
    author = "L{\"u}scher, M.",
    title = "{Volume Dependence of the Energy Spectrum in Massive Quantum Field Theories. 2. Scattering States}",
    reportNumber = "DESY-86-034",
    doi = "10.1007/BF01211097",
    journal = "Commun. Math. Phys.",
    volume = "105",
    pages = "153--188",
    year = "1986"
}

@article{Maas:2017xzh,
    author = {Maas, Axel and Sondenheimer, Ren{\'e} and T{\"o}rek, Pascal},
    title = "{On the observable spectrum of theories with a Brout{\textendash}Englert{\textendash}Higgs effect}",
    eprint = "1709.07477",
    archivePrefix = "arXiv",
    primaryClass = "hep-ph",
    doi = "10.1016/j.aop.2019.01.010",
    journal = "Annals Phys.",
    volume = "402",
    pages = "18--44",
    year = "2019"
}

@article{Maas:2014pba,
    author = "Maas, Axel and Mufti, Tajdar",
    title = "{Spectroscopic analysis of the phase diagram of Yang-Mills-Higgs theory}",
    eprint = "1412.6440",
    archivePrefix = "arXiv",
    primaryClass = "hep-lat",
    doi = "10.1103/PhysRevD.91.113011",
    journal = "Phys. Rev. D",
    volume = "91",
    number = "11",
    pages = "113011",
    year = "2015"
}

@article{Kulkarni:2022bvh,
    author = "Kulkarni, Suchita and Maas, Axel and Mee, Se{\'a}n and Nikolic, Marco and Pradler, Josef and Zierler, Fabian",
    title = "{Low-energy effective description of dark $Sp(4)$ theories}",
    eprint = "2202.05191",
    archivePrefix = "arXiv",
    primaryClass = "hep-ph",
    doi = "10.21468/SciPostPhys.14.3.044",
    journal = "SciPost Phys.",
    volume = "14",
    number = "3",
    pages = "044",
    year = "2023"
}

@article{Kogut:2000ek,
    author = "Kogut, J. B. and Stephanov, Misha A. and Toublan, D. and Verbaarschot, J. J. M. and Zhitnitsky, A.",
    title = "{QCD - like theories at finite baryon density}",
    eprint = "hep-ph/0001171",
    archivePrefix = "arXiv",
    reportNumber = "SUNY-NTG-00-11",
    doi = "10.1016/S0550-3213(00)00242-X",
    journal = "Nucl. Phys. B",
    volume = "582",
    pages = "477--513",
    year = "2000"
}

@article{Bennett:2019jzz,
    author = "Bennett, Ed and Hong, Deog Ki and Lee, Jong-Wan and Lin, C. -J. David and Lucini, Biagio and Piai, Maurizio and Vadacchino, Davide",
    title = "{Sp(4) gauge theories on the lattice: $N_f=2$ dynamical fundamental fermions}",
    eprint = "1909.12662",
    archivePrefix = "arXiv",
    primaryClass = "hep-lat",
    doi = "10.1007/JHEP12(2019)053",
    journal = "JHEP",
    volume = "12",
    pages = "053",
    year = "2019"
}

@article{Dengler:2024maq,
    author = "Dengler, Yannick and Maas, Axel and Zierler, Fabian",
    title = "{Scattering of dark pions in Sp(4) gauge theory}",
    eprint = "2405.06506",
    archivePrefix = "arXiv",
    primaryClass = "hep-lat",
    doi = "10.1103/PhysRevD.110.054513",
    journal = "Phys. Rev. D",
    volume = "110",
    number = "5",
    pages = "054513",
    year = "2024"
}

@article{Martins:2024ilw,
    author = "Martins, Sofie and Bowes, Laurence Sebastian and Drach, Vincent and Fritzsch, Patrick and Rago, Antonio and Romero-L{\'o}pez, Fernando",
    title = "{Determination of the pseudoscalar decay constant from SU(2) with two fundamental flavors}",
    eprint = "2412.06471",
    archivePrefix = "arXiv",
    primaryClass = "hep-lat",
    reportNumber = "MIT-CTP/5765",
    doi = "10.22323/1.466.0149",
    journal = "PoS",
    volume = "LATTICE2024",
    pages = "149",
    year = "2025"
}

@article{Hernandez:2019qed,
    author = "Hern{\'a}ndez, Pilar and Pena, Carlos and Romero-L{\'o}pez, Fernando",
    title = "{Large $N_c$ scaling of meson masses and decay constants}",
    eprint = "1907.11511",
    archivePrefix = "arXiv",
    primaryClass = "hep-lat",
    doi = "10.1140/epjc/s10052-019-7395-y",
    journal = "Eur. Phys. J. C",
    volume = "79",
    number = "10",
    pages = "865",
    year = "2019"
}

@article{10.1145/3711897,
author = {Fjelde, Tor Erlend and Xu, Kai and Widmann, David and Tarek, Mohamed and Pfiffer, Cameron and Trapp, Martin and Axen, Seth D. and Sun, Xianda and Hauru, Markus and Yong, Penelope and Tebbutt, Will and Ghahramani, Zoubin and Ge, Hong},
title = {Turing.jl: a general-purpose probabilistic programming language},
year = {2025},
publisher = {Association for Computing Machinery},
address = {New York, NY, USA},
url = {https://doi.org/10.1145/3711897},
doi = {10.1145/3711897},
journal = {ACM Trans. Probab. Mach. Learn.},
month = feb,
}

@InProceedings{pmlr-v84-ge18b,
  title = 	 {Turing: A Language for Flexible Probabilistic Inference},
  author = 	 {Ge, Hong and Xu, Kai and Ghahramani, Zoubin},
  booktitle = 	 {Proceedings of the Twenty-First International Conference on Artificial Intelligence and Statistics},
  pages = 	 {1682--1690},
  year = 	 {2018},
  editor = 	 {Storkey, Amos and Perez-Cruz, Fernando},
  volume = 	 {84},
  series = 	 {Proceedings of Machine Learning Research},
  month = 	 {09--11 Apr},
  publisher =    {PMLR},
  url = 	 {https://proceedings.mlr.press/v84/ge18b.html},
}

@article{Ramos:2018vgu,
    author = "Ramos, Alberto",
    title = "{Automatic differentiation for error analysis of Monte Carlo data}",
    eprint = "1809.01289",
    archivePrefix = "arXiv",
    primaryClass = "hep-lat",
    doi = "10.1016/j.cpc.2018.12.020",
    journal = "Comput. Phys. Commun.",
    volume = "238",
    pages = "19--35",
    year = "2019"
}

@article{Cucchieri:1995pn,
    author = "Cucchieri, Attilio and Mendes, Tereza",
    title = "{Critical slowing down in SU(2) Landau gauge fixing algorithms}",
    eprint = "hep-lat/9511020",
    archivePrefix = "arXiv",
    reportNumber = "NYU-TH-95-11-02",
    doi = "10.1016/0550-3213(96)00177-0",
    journal = "Nucl. Phys. B",
    volume = "471",
    pages = "263--292",
    year = "1996"
}

@article{Joswig:2022qfe,
    author = "Joswig, Fabian and Kuberski, Simon and Kuhlmann, Justus T. and Neuendorf, Jan",
    title = "{pyerrors: A python framework for error analysis of Monte Carlo data}",
    eprint = "2209.14371",
    archivePrefix = "arXiv",
    primaryClass = "hep-lat",
    reportNumber = "MITP-22-052, MS-TP-22-31",
    doi = "10.1016/j.cpc.2023.108750",
    journal = "Comput. Phys. Commun.",
    volume = "288",
    pages = "108750",
    year = "2023"
}

@article{Wolff:2003sm,
    author = "Wolff, Ulli",
    collaboration = "ALPHA",
    title = "{Monte Carlo errors with less errors}",
    eprint = "hep-lat/0306017",
    archivePrefix = "arXiv",
    reportNumber = "HU-EP-03-32, SFB-CPP-03-12",
    doi = "10.1016/S0010-4655(03)00467-3",
    journal = "Comput. Phys. Commun.",
    volume = "156",
    pages = "143--153",
    year = "2004",
    note = "[Erratum: Comput.Phys.Commun. 176, 383 (2007)]"
}

@article{Schaefer:2010hu,
    author = "Schaefer, Stefan and Sommer, Rainer and Virotta, Francesco",
    collaboration = "ALPHA",
    title = "{Critical slowing down and error analysis in lattice QCD simulations}",
    eprint = "1009.5228",
    archivePrefix = "arXiv",
    primaryClass = "hep-lat",
    reportNumber = "DESY-10-151, SFB-CPP-10-81, HU-EP-10-55",
    doi = "10.1016/j.nuclphysb.2010.11.020",
    journal = "Nucl. Phys. B",
    volume = "845",
    pages = "93--119",
    year = "2011"
}

@article{Philipsen:1996af,
    author = "Philipsen, O. and Teper, M. and Wittig, H.",
    title = "{On the mass spectrum of the SU(2) Higgs model in (2+1)- dimensions}",
    eprint = "hep-lat/9602006",
    archivePrefix = "arXiv",
    reportNumber = "OUTP-95-40-P, DESY-96-017",
    doi = "10.1016/0550-3213(96)00156-3",
    journal = "Nucl. Phys. B",
    volume = "469",
    pages = "445--472",
    year = "1996"
}

@article{APE:1987ehd,
    author = "Albanese, M. and others",
    collaboration = "APE",
    title = "{Glueball Masses and String Tension in Lattice QCD}",
    reportNumber = "ROM2F-87-014",
    doi = "10.1016/0370-2693(87)91160-9",
    journal = "Phys. Lett. B",
    volume = "192",
    pages = "163--169",
    year = "1987"
}

\end{document}